# Neighbor List Collision-Driven Molecular Dynamics Simulation for Nonspherical Hard Particles.
# I. Algorithmic Details

Aleksandar Donev,[1,2] Salvatore Torquato,[1,2,3,*] and Frank H. Stillinger[3]

[1]*Program in Applied and Computational Mathematics,*
*Princeton University, Princeton NJ 08544*

[2]*Materials Institute, Princeton University, Princeton NJ 08544*

[3]*Department of Chemistry, Princeton University, Princeton NJ 08544*

## Abstract

In this first part of a series of two papers, we present in considerable detail a collision-driven molecular dynamics algorithm for a system of nonspherical particles, within a parallelepiped simulation domain, under both periodic or hard-wall boundary conditions. The algorithm extends previous event-driven molecular dynamics algorithms for spheres, and is most efficient when applied to systems of particles with relatively small aspect ratios and with small variations in size. We present a novel partial-update near-neighbor list (NNL) algorithm that is superior to previous algorithms at high densities, without compromising the correctness of the algorithm. This efficiency of the algorithm is further increased for systems of very aspherical particles by using bounding sphere complexes (BSC). These techniques will be useful in any particle-based simulation, including Monte Carlo and time-driven molecular dynamics. Additionally, we allow for a nonvanishing rate of deformation of the boundary, which can be used to model macroscopic strain and also alleviate boundary effects for small systems. In the second part of this series of papers we specialize the algorithm to systems of ellipses and ellipsoids and present performance results for our implementation, demonstrating the practical utility of the algorithm.

---

[*] Electronic address: `torquato@electron.princeton.edu`

## I. INTRODUCTION

Classical molecular dynamics (MD) simulations have been used to study the properties of particle systems for many decades. The first MD studies used the simplest multi-particle system, the hard-sphere fluid/solid [1], which is very rich in behavior. Subsequently, methods were developed to follow the dynamics of a system of soft spheres, i.e., particles interacting with a spherically symmetric and continuous interparticle potential, usually with a hard cutoff on the range of the interaction. The algorithms needed to simulate the two types of systems are rather different, and difficult to combine [12]. For soft particles, one needs to integrate a system of ordinary differential equations given by Newton's law of motion. However, for hard particles the interaction potential is singular and the task of integrating the equations of motion becomes a problem of processing a sequence of binary collisions between the particles[1], or collisions of the particles with the hard walls of a container, if any. In other words, for hard particles, one needs to predict and process a sequence of discrete *events* of vanishing duration.

The algorithm for hard particles therefore becomes *event-driven*, as opposed to the *time-driven* algorithm for soft-particle MD in which time changes in small steps and the equations of motion are integrated. Event-driven algorithms have the task of scheduling a sequence of events predicted to happen in the future. The simulation is advanced to the time of the event with the smallest scheduled time (the *impending event*) and that event is processed. The schedule of events is updated if necessary and the same process is repeated. In molecular dynamics, the primary kind of event are binary collisions, so the simulation becomes *collision-driven*. This kind of collision-driven approach was used for the very first MD simulation of the hard-disk system [1], and has since been extended and improved in a variety of ways, most importantly, to increase the efficiency of the algorithm. All of these improvements, namely, delayed particle updates, the cell method, the use of a *heap* for the event queue, etc., appear in almost any efficient hard-particle MD algorithm. Systems of many thousands of hard disks or spheres can be studied on a modern personal computer using such algorithms, and in the past decade the method has been extended to handle more complex simulations, such as particles in a velocity field [30].

---

[1] Multi-particle collisions have zero probability of occurring and will not be considered here.

However, we are aware of only one collision-driven simulation in the literature for non-spherical particles, namely, one for very thin rods (needles) [8]. Other molecular dynamics simulations for hard nonspherical particles have used a time-driven approach [2], which is simpler to implement than the event-driven approach but inferior in both accuracy and efficiency (at high densities). Two kinds of smooth shapes are used frequently to model aspherical particles: *spherocylinders* (a cylinder with two spherical caps), and *ellipsoids*. Both can become spheres in a suitable limit. Spherocylinders are analytically much simpler then ellipsoids; however, they are always axisymmetric and cannot be oblate.

The primary reason event-driven algorithms have not yet been used for nonspherical particles is that a high-accuracy collision-driven scheme for nontrivial particle shapes and sufficiently large systems is very demanding. Computers have only recently reached the necessary speeds, and a proper implementation involves a significant level of code complexity. In this paper, we present in detail a collision-driven molecular dynamics algorithm for a system of hard nonspherical particles. The algorithm is based on previous event-driven MD approaches for spheres, and in particular the algorithms of Lubachevsky [19] and Sigurgeirsson *et al.* [2]. While in principle the algorithm is applicable to any particle shape, we have specifically tailored it for smooth particles for which it is possible to introduce and easily evaluate continuously differentiable *overlap potentials*. Additionally, it is assumed that the particles have a spherically symmetric moment of inertia, so that in-between collisions their angular velocities are constant. Furthermore, the algorithm is most efficient when applied to relatively dense and homogeneous systems of particles which do not differ widely in size (i.e., the degree of polydispersity is small). We focus in this work on systems with *lattice-based* boundaries, under both periodic or hard-wall boundary conditions. The main innovations and strengths of the proposed algorithm are:

- It specifically allows for non-axially symmetric particles by using quaternions in the representation of orientational degrees of freedom, unlike previous hard-particle algorithms which have been restricted just to needles, spherocylinders or spheroids;

- The particle-shape-dependent components of the algorithm are clearly separated from general concepts, so that it is (at least in principle) easy to adapt the algorithm to different particle shapes. In the second part of this series of papers we present in detail the implementation of these components for ellipses and ellipsoids;

- It explicitly allows for time-dependent particle shapes and for time-dependent shape of the boundary cell, which enables a range of nonequilibrium applications and also Parinello-Rahman-like [26] constant-pressure molecular dynamics;

- It corrects some assumptions in traditional hard-sphere algorithms that are not correct for nonspherical particles or when boundary deformation is included, such as the *nearest image convention* in periodic systems and the claim that there must be an intervening collision between successive collisions of a given pair of particles;

- It is the first rigorous event-driven MD algorithm to incorporate *near-neighbor lists*, by using the concept of *bounding neighborhoods*. This is a very significant improvement for very aspherical particles and/or at high densities, and has some advantages over the traditional *cell method* even for hard spheres because it allows a close monitoring of the collision history of the algorithm;

- It is the first algorithm to specifically address the problem of efficient near-neighbor search for very elongated or very flat particles by introducing the concept of *bounding sphere complexes*. The algorithm also clearly separates neighbor-search in a static environment (where particle positions are fixed) from its use in a dynamic environment (where particles move continuously), thus enabling one to easily incorporate additional neighbor search techniques. We emphasize that the developed near neighbor search techniques will improve all particle-based simulations, including Monte Carlo and time-driven molecular dynamics; and

- It is documented in detail with pseudo-codes which closely follow the actual Fortran 95 code used to implement it for ellipses and ellipsoids.

One motivation for developing this algorithm has been to extend the Lubachevsky-Stillinger sphere-packing algorithm [20, 21] to nonspherical particles. We have successfully used our implementation to obtain many interesting results for random and ordered packings of ellipses and ellipsoids [6]. The algorithm can also be used to study equilibrium properties of hard-ellipse and hard-ellipsoid systems, and we give several illustrative applications in the second paper of this series. In the second paper we also numerically demonstrate that our novel neighbor-search techniques can speed the simulation by as much as two orders of magnitude or more at high densities and/or for very aspherical particles, as compared to direct adaptations of traditional hard-sphere schemes.

We begin by presenting preliminary information and the basic ideas behind the algorithm in Section. II. We then focus on the important task of improving the efficiency of the algorithm by focusing on neighbor search in Section III, and present both the classical cell method and our adaptation of near-neighbor lists. Detailed pseudocodes for all major steps in the algorithm for general nonspherical particles are given in Section IV, and these are continued in the second part of this series of papers for the specific case of ellipses and ellipsoids.

## II. PRELIMINARIES

In this section we give some background information and a preliminary description of the algorithm. First, we discuss the impact the shape of the particles has on the algorithm. Then we briefly describe the two main approaches to hard-particle molecular dynamics, time-driven and event-driven. Finally, we discuss boundary conditions in our event-driven molecular dynamics algorithm and also the possibility of performing event-driven MD in different ensembles. Bold symbols are reserved for vectors and matrices, and subscripts are used to denote their components. Subscripts or superscripts are used heavily to add specificity to various quantities, for example, $\mathbf{r}$ denotes position, while $\mathbf{r}_A$ denotes the position of some particle $A$. We denote the numerical precision with $\epsilon \ll 1$, and use subscripted $\epsilon$'s for various user-set (small) numerical tolerances. We often omit explicit functional dependencies when they are clearly implied by the context and it is not important to emphasize them, for example, $f$ and $f(t)$ will be used interchangeably.

### A. Particle Shape

We consider a system of $N$ hard particles whose only interactions are given by impenetrability constraints, although it is easy to allow for additional external fields which are independent of the particles (such as gravity). Many of the techniques developed here are also used to deal with particles interacting with a soft potential if there is a hard cutoff on the potential. We discuss the special case of orientation-less particles, namely spheres, at length, and we use hard ellipsoids to illustrate the extensions to nonspherical particles. We will use the terms sphere and ellipsoid in any dimension, but sometimes we will be more specific and distinguish between disk and ellipse in two dimensions, and sphere and ellipsoid in three dimensions.

Spheres are a very important special case not only because of their simplicity, but also because *bounding spheres* are a necessary ingredient when dealing with aspherical particles. A bounding sphere for a particle is centered at the *centroid* of the particle and has the minimal possible diameter $D_{\max} = 2O_{\max}$ so that it fully encloses the particle itself. Here by centroid we mean a geometrically special point chosen so that the bounding sphere is as small as possible (i.e., it should be chosen to be as close as possible to the midpoint of the longest line segment joining two points of the particle). For example, for an ellipsoid, the bounding sphere has the same center as the ellipsoid and its diameter is equal to the largest axes of the ellipsoid. The importance of bounding spheres is that they provide a quick and analytically simple way to test for overlap of two particles: Two particles cannot overlap if their bounding spheres do not overlap. Occasionally we make use of *contained spheres*, which are also centered at the centroid of the particle and have the maximal possible diameter $D_{\min} = 2O_{\min}$ so that they are fully within the particle itself. For ellipsoids their diameter is the smallest axes. Note that two particles must overlap if their contained spheres overlap. The efficiency of the EDMD algorithm described in this paper is primarily determined by the *aspect ratio* $\alpha = D_{\max}/D_{\min}$. The greater the deviation of $\alpha$ from unity, the worse the efficiency because the bounding/contained spheres become worse approximations for the particles and because of the increasing importance of particle orientations. We propose a novel *near-neighbor list* technique for dealing with very aspherical particles, in addition to the standard *cell method*.

### B.  Molecular Dynamics

The goal of our algorithm is to simulate the motion of the particles in time as efficiently as possible, while taking into account the interactions between the particles. For hard-particle systems, the only interactions occur during *binary collisions* of the particles. The goal of hard-particle *molecular dynamics* (MD) algorithms is to correctly predict the time-ordered sequence of particle collisions. Additionally, there may be obstacles such as hard walls with which the particles can collide. Next we briefly introduce the main ideas behind the two main approaches to hard-particle MD, time-driven and event-driven MD. This preliminary presentation will be helpful in understanding the rest of this section. Further details on the event-driven algorithm are given in Section IV.

*1. Time Driven MD*

The *Time-Driven Molecular Dynamics* (TDMD) approach is inspired by MD simulations of systems of soft particles (i.e., particles interacting with a continuous interaction potential). It has been adapted also to the simulation of hard-particle systems, particularly nonspherical particles [2, 28], mainly because of its simplicity. In this approach, all of the particles are displaced *synchronously* in small *time steps* $\Delta t$ and a check for overlap between the particles is done. If any two particles overlap, time is rolled back until the approximate moment of initial overlap, i.e., the time of collision, and the collision of the particles is processed (i.e., the momenta of the colliding particles are updated), and the simulation continued. The main disadvantage of this approach is that it is *not* rigorous, in the sense that collisions may be missed or the correct ordering of a sequence of successive collisions may be mis-predicted (particularly in dense systems). To ensure a reasonably correct prediction of the system dynamics, a very small time step must be used and this is inefficient. Nonetheless, since only checking for overlap between particles is needed, the simplicity of the method is a very attractive feature. Additionally, such an approach is parallelizable with the same techniques as any other MD algorithm (for example, domain decomposition).

*2. Event Driven MD*

An alternative *rigorous* approach is to use *Event-Driven Molecular Dynamics* (EDMD), based on a rather general model of discrete event-driven simulation. In EDMD, instead of advancing time independently of the particles as in TDMD, time is advanced from one *event* to the next event, where an event is a binary particle collision, or a collision of a particle with an obstacle (hard wall). Other types of events will be discussed shortly, however, collisions are the central type of event so we label the approach is more specifically *Collision-Driven Molecular Dynamics* (CDMD). We will however continue to use the abbreviation EDMD since the term event-driven is widely used in the literature.

Efficient implementations of EDMD are *asynchronous*: each particle is at the point in time when the last event involving it happened. Each particle predicts what its *impending event* is and when it is expected to happen. All of these events are entered into a priority *event queue* (typically implemented by a heap), which allows for quick extraction of the next event to happen. The positions and momenta of the particles involved in this event are updated, the particles' next event predicted, the event queue updated, and the simulation continued with the next event. Sometimes events may be mis-predicted. For example, a particle $i$ may predict a collision with particle $j$, but another (third party) particle $m$ may collide with $j$ before $i$ has time to. A special event called a *check* needs to be introduced, and it amounts to simply (re-)predicting the impending event for a given particle. Given infinite numerical precision, this kind of approach rigorously follows the dynamics of the system.

The computationally expensive step in EDMD is the prediction of the impending event of a given particle $i$ (even though asymptotically the event-queue operations dominate). This typically involves the expensive (especially for nonspherical particles) step of predicting the time of collision between the particle $i$ and a set of other particles $j$. In the simplest approach, one would predict the time of collision between $i$ and all other particles and choose the smallest one, but a much more efficient approach is described in Section III. For spheres moving along straight lines, predicting the time of collision merely amounts to finding the first positive root (if any) of a quadratic equation, and is very fast. Therefore, for spherical particles EDMD always outperforms TDMD by orders of magnitude, *and* it is rigorous. For nonspherical particles, collision predictions are much more involved, but for algebraically simple smooth particle shapes it is expected that EDMD will still outperform TDMD for a wide range of densities. Furthermore, there are systems for which TDMD is not possible, and one must use EDMD, such as systems of hard line segments [8]. Note however that the efficiency of the EDMD approach is possible only because the motion of the particles between events can be predicted *a priori*, and because binary collisions only affect the two colliding particles. In cases when these assumptions are not true, TDMD may be the only option. Additionally, it is very important to note that the EDMD algorithm is inherently non-parallelizable due its sequential processing of the events. Some attempts have been made to parallelize the method [18] by using the locality of the interactions, and very recently actual implementations have appeared [24]. We will defer any discussion of parallelization to future publications.

## C. Boundary Conditions

In this paper we consider MD in a simple bounded simulation domain embedded in a Euclidean space $E^d$ of dimensionality $d$. In particular, we focus exclusively on *lattice-based boundaries*. This means that the simulation domain, which the particles never leave, is a *parallelepiped* defined by $d$ *lattice vectors*, $\boldsymbol{\lambda}_1, \ldots, \boldsymbol{\lambda}_d$. The simulation domain, or *unit cell*, is a collection of points with $d$ *relative coordinates* $\mathbf{r}$ in the interval $[0, 1]$, and corresponding Cartesian coordinates

$$\mathbf{r}^{(E)} = \sum_{k=1}^{d} r_k \boldsymbol{\lambda}_k = \boldsymbol{\Lambda} \mathbf{r}, \tag{1}$$

where $\boldsymbol{\Lambda}$ is a square invertible matrix representing the *lattice*, and contains the lattice vectors as columns. The volume of the unit cell is given by the positive determinant $|\boldsymbol{\Lambda}|$. As illustrated in Fig. 1(a), the parameters describing the geometry of a lattice-based boundary separate into components along different "dimensions", meaning along different lattice vectors. For example, there are $d$ perpendicular distances $L_k$ between the "left" and "right" faces of the parallelepiped along $\boldsymbol{\lambda}_k$ (meaning the two $(d-1)$-dimensional faces spanned by all lattice vectors other than $\boldsymbol{\lambda}_k$), one for each dimension $k = 1, \ldots, d$. We assume that in three dimensions the lattice vectors form a right-handed coordinate system, so each lattice vector can be identified as defining the $x$ ($k=1$), $y$ ($k=2$), or $z$ ($k=3$) axis.

Additionally, we allow either *periodic* or *hard-wall* boundary conditions (BCs) to be specified *independently* along each dimension, that is, the "left" and "right" faces of the unit cell along each dimension can either be hard walls or periodic boundaries. The most commonly used BCs are fully periodic, and one can interpret periodic systems as being on a topological torus whose distance geometry is determined by the metric tensor $\mathbf{G} = \boldsymbol{\Lambda}^T \boldsymbol{\Lambda}$ (a "flat" torus), or one can interpret periodic systems as being infinite and covering all of Euclidean space with identical copies of the unit cell and the particles in this unit cell. We will refer to the particles in the unit cell as *original particles* and simply identify them with an integer $i = 1, \ldots, N$. There are infinitely many *image* or *virtual particles* for every original particle, translated from the original by an integer number of lattice vectors $\mathbf{n}_c \in \mathcal{Z}$. We identify such an image of particle $i$ with a pair of integers $(i, v)$, where the *image* or *virtual* *identifier* $v$ denotes the particular image in question, $\mathbf{n}_c = \mathbf{n}_c(v)$. Traditional hard-sphere algorithms have used the so-called nearest image convention, which assumes that only one (easy-to-identify) image of a given particle $j$ can overlap particle $i$ and thus $\mathbf{n}_c$ needs not be explicitly stored. However, this assumption is wrong for nonspherical particles, where several images of a given particle $j$ can overlap an original particle $i$.

One almost never needs to worry about any but the $3^d$ (9 in two, and 27 in three dimensions) images of the unit cell that neighbor the original one (first neighbors, including the cell under consideration). We number these images with $v = -(3^d - 1)/2, ..., (3^d - 1)/2$, so that images with opposite $\mathbf{n}_c$'s have identifiers of equal magnitude but opposite sign, $\mathbf{n}_c(-v) = -\mathbf{n}_c(v)$, as illustrated in two dimensions in Fig. 1(b). Note that $(i, 0) \equiv i$. When considering *ordered* pairs of particles $[i, (j, v)]$, one of the particles, $i$, is always original, while the other one, $(j, v)$, can be an image or an original. Due to our choice of image-numbering, we can alternatively consider every such *unordered* pair to be composed of particles $j$ and $(i, -v)$, $\{i, (j, v)\} \equiv \{j, (i, -v)\}$.

### 1. Boundary Deformation

In our simulations, we use *relative* coordinates $\mathbf{r}$ (i.e., expressed in terms of the lattice vectors), and likewise relative velocities $\mathbf{v}$, and convert these into their *Euclidian* representations $\mathbf{r}^{(E)}$ and $\mathbf{v}^{(E)}$ when necessary. The relative position of an image particle is $\mathbf{r} + \mathbf{n}_c$. The conversion between the two representations adds computational overhead due the matrix-vector multiplication[2] in (1). Some of this overhead can be avoided by only using Euclidean positions, however, we have chosen to express all positions relative to the lattice. The primary reason for this choice is that we allow the lattice to deform, that is, we allow for a *lattice velocity* $\dot{\mathbf{\Lambda}}$. In our algorithms, the lattice can change *linearly* with time,

$$\Delta \mathbf{\Lambda} = \dot{\mathbf{\Lambda}} \Delta t,$$

even though a more correct approach is to have a constant *strain rate*

$$\dot{\boldsymbol{\epsilon}} = \dot{\mathbf{\Lambda}} \mathbf{\Lambda}^{-1}, \tag{2}$$

---

[2] Note however that when the lattice is $\mathbf{\Lambda} = \mathbf{I}$, which is the usual choice unless a special unit cell is needed, all the matrix-vector multiplications become trivial. Our implementation has a special mode for this simple but important case.

that is, to have an exponential time evolution of the lattice,

$$\mathbf{\Lambda}(t) = \exp\left(\dot{\boldsymbol{\epsilon}}t\right) \mathbf{\Lambda}(0).$$

The identification of $\boldsymbol{\epsilon} = (\Delta\mathbf{\Lambda})\mathbf{\Lambda}^{-1}$ with the macroscopic strain is explained in Ref. [31], and we choose it to be symmetric, $\dot{\boldsymbol{\epsilon}} = \dot{\boldsymbol{\epsilon}}^T$, to eliminate rotations of the unit cell.

In our approach, since the positions of the particles are relative to the lattice, the particles move together with the lattice. This is necessary in order to simulate isotropic systems. Namely, had the positions of the particles been independent of the lattice and the lattice deformed, for example, uniformly contracted, the image particles would move with the lattice, but the originals would not, and this would lead to artificial effects at the boundary of the unit cell. However, using relative positions is not without a cost. Consider a particle at relative position $\mathbf{r}$ which moves with constant relative velocity $\mathbf{v}$. It's Euclidean position is a parabola,

$$\mathbf{r}^{(E)}(t) = \left(\mathbf{\Lambda} + \dot{\mathbf{\Lambda}}t\right)(\mathbf{r} + \mathbf{v}t) = \mathbf{\Lambda}\mathbf{r} + \left(\mathbf{\Lambda}\mathbf{v} + \dot{\mathbf{\Lambda}}\mathbf{r}\right)t + \left(2\dot{\mathbf{\Lambda}}\mathbf{v}\right)\frac{t^2}{2}, \qquad (3)$$

rather than a straight line. We can identify the instantaneous Euclidean position and velocity as well as the acceleration to be

$$\mathbf{r}^{(E)} = \mathbf{\Lambda}\mathbf{r} \qquad (4)$$

$$\mathbf{v}^{(E)} = \mathbf{\Lambda}\mathbf{v} + \dot{\mathbf{\Lambda}}\mathbf{r} \qquad (5)$$

$$\mathbf{a}^{(E)} = 2\dot{\mathbf{\Lambda}}\mathbf{v}. \qquad (6)$$

This complicates, for example, the calculation of the time of collision of two moving spherical particles. Ordinarily a *quadratic* equation needs to be solved, but when the lattice deforms, a *quartic* equation needs to be solved instead. To our knowledge, our algorithm is the first EDMD algorithm to include a deforming boundary.

In our algorithm, the lattice velocity is an *externally* imposed quantity, and our goal is to simulate the motion of the particles as the boundary deforms, for example, in order to study shear banding in systems of ellipsoids [10]. It is the usual case that the boundary deforms slowly compared to the motion of the particles. As the boundary deforms, the unit cell becomes less and less orthogonal, and so in long-time simulations some form of orthogonalization of the unit cell might be necessary (we have not experimented with such techniques). Previously used constant-shear MD techniques [17] do not have this problem, however, they are also not capable or simulating arbitrary shears and are plagued with boundary effects.

Unlike in TDMD, in EDMD it is *not* possible to couple the motion of the boundary to all of the particles, as is done in Parinello-Rahman MD [26]. This is because the efficiency of the method depends critically on the fact that particles move independently between collisions and that particle collisions only affect the colliding particles. However, a pseudo-PRMD approach is possible, in which the lattice velocity is updated after a certain number of particle collisions, and the simulation is essentially restarted with a new lattice velocity. We have some preliminary positive experience with such a method. Unit cell dynamics is also needed to properly study anisotropic liquids, which is very important for nonspherical particles [15]. A deforming boundary can also be used to model macroscopic strain in multiscale simulations (for example, to simulate granular flow) in which microscale MD is used to obtain material properties needed for a macroscopic continuum simulation.

## D. EDMD in Different Ensembles

Molecular dynamics is often performed in ensembles different from the $NVE$ one, and in particular, constant temperature and constant pressure are often desired. For this purpose, various thermostats have been developed. However, these are usually designed to be used with time-driven MD and systems of soft particles. We are in fact aware of no work that explicitly discusses thermostats for event-driven MD.

Hard particle systems are inherently athermal due to the lack of energy scale, and the pressure and time scaling are therefore arbitrary. Sophisticated temperature or pressure thermostats are thus not usually needed. In particular, simple *velocity rescaling* can be used to keep the temperature at the desired value. The average translational kinetic energy $E_k$ per particle can be calculated and then *both* the translational and angular velocities scaled by the factor $s = \sqrt{dkT/2E_k}$ and the simulation essentially restarted[3]. This kind of temperature control is needed when, for example, the particles grow or shrink in size, since this leads to nonconservative collision dynamics and an overall heating or cooling of the system. Although velocity rescaling is simple and convenient, it has serious defficiencies

---

[3] Equipartition of energy is usually maintained by the collision dynamics, and therefore we usually do not use a different scaling for the angular velocities.

[11], and a true canonical thermostat may be needed in some applications. For this purpose, an Andersen thermostat [9] can be included in collision-driven algorithms by considering the thermostat as a possible collision partner (with an appropriate Poisson distribution of collision times). We do not include such a stochastic temperature thermostat in our algorithm explicitly, since we have not used it.

A Parinello-Rahman-like isostress (isopressure) thermostat [26] cannot be directly included in collision-driven algorithms, since it implicitly couples the motion of all particles via the deformation of the unit cell and thus destroys the asynchronous efficiency of collision-driven approaches. Such a thermostat is often needed, even for hard particles, in simulations of crystal phases in order to keep the internal stress tensor isotropic and to allow for changes of the crystal unit cell. We have used constant shear boundary deformations, as described in Section II C 1, to implement a partial isostress thermostat in which the shear rate is periodically updated to reflect the asymmetry of the stress tensor. This approach has had a mixed success and additional work is required to improve it, especially for anisotropic systems of (aspherical) particles [15].

### III. SPEEDING UP THE SEARCH FOR NEIGHBORS

Identifying the near-neighbors of a given particle has the most important impact on efficiency in almost all simulations of particle systems, particularly when the interparticle interactions are short-range. In both MC and TDMD algorithms it is important to quickly identify only the particles that are within the interaction cutoff distance $l_{\text{cutoff}}$ (here distance is measured in the metric appropriate for the interaction) from a given particle and only evaluate the force or interaction energy with these particles. Since the number of such near neighbors is typically a small constant (of the order of $5 - 20$, strongly increasing with increasing dimensionality $d$), this ensures that the computational effort needed to evaluate the forces on the particles or potential energy scales *linearly* with the number of particles $N$, as opposed to the quadratic complexity of checking all pairs of particles. In EDMD algorithms, it is useless to predict collisions between all pairs of particles since only nearby particles are actually likely to collide, and in fact $N \log N$ scaling can be obtained in EDMD by only predicting collisions between a given particle and a bounded (small) number of near neighbors.

In this section we describe the traditional cell method for speeding up neighbor search, and its implementation for lattice-based boundaries. We then propose a novel method based on the familiar concept of near-neighbor lists, which offers significant computational savings over the pure cell method for very aspherical particles, as demonstrated numerically in the second part of this series of papers.

### A. The Cell Method

One traditional method for neighbor search in particle systems is the so-called *cell method* (see for example Ref. [3]). It consists of partitioning the simulation domain into $N_c$ disjoint cells and maintaining for each cell a list of all the particles whose centroids are within it. Then, for a given particle $i$, only the particles $j$ in the neighboring cells (including periodic images of cells) of $i$'s cell are considered neighbors of particle $i$. The shape of the cells can be chosen arbitrarily, so long as the union of all cells covers the whole simulation domain, and so long as for any given cell $c$ one can (easily) identify all *neighbor cells* $c_n$ that contain a point within Euclidean distance $l_{\text{cutoff}}$ from a point in $c$. This enables a *rigorous* identification of all particles whose centroids are within a given cutoff distance from the centroid of a given particle. The essential aspect of the cell method is that the partitioning into cells is independent of the motion of the particles, so that even as the particles move one can continue to rely on using the cells to rigorously identify neighbors in constant time. This is a unique and necessary strength of the cell method, and *all* of our simulations use the cell method in some form, to ensure correctness while maintaining efficiency.

For maximal efficiency, it seems that it is best to choose the cells as small as possible, but ensuring that only cells which actually share a boundary (i.e., are adjacent) need to be considered as neighbors. While this is obvious for MC or TDMD simulations, it is not so obvious for EDMD. For event-driven algorithms, it can be theoretically predicted and verified computationally that it is best to choose the number of cells to be of the order of the number of particles [30], and computational experiments suggest that there should be about one particle per cell. For moderately to very dense systems, one should therefore choose the cells so that the maximal Euclidean distance between two points in the same cell, $L_c$, is as close to the largest enclosing sphere diameter $D_{\max}$ as possible,

$$L_c = (1 + \epsilon_L) D_{\max}.$$

We have verified this choice to be optimal in our extensive computational experience and consistently try to maximize the number of cells in all our simulations.

We note that in some simulations the shape of the particles changes. For example, the Lubachevsky-Stillinger algorithm [20] generates dense packings of particles by performing an EDMD simulation while the particles expand uniformly (for example, for spheres the radius $O$ changes linearly with time with a constant expansion rate $\gamma$, $O(t) = O + \gamma t$). In such cases one must ensure that a sphere of diameter $D_{\max}$ can always enclose any of the particles in the system. A suitable value for $D_{\max}$ can be found, for example, by assuming that the final packing is the densest possible (or fills space completely if an exact result for the maximal density is not known). It is also important to note that it is sometimes needed to find all particles whose centroids are within a distance larger then $D_{\max}$ from the centroid of a given particle. This is not a problem for the cell method, as one can simply include as many additional cells in the search as needed to guarantee that the search is rigorous. For example, one may need to include neighbor cells of the neighbor cells (i.e., second-neighbor cells).

The need to adjust the partitioning of Euclidian space into cells to the shape of the simulation domain is the most difficult aspect of using the cell-method. Most simulations in the literature have been done with spherical particles and in cubic simulation domains, and the partitioning of the simulation domain is a simple Cartesian grid (mesh) of cells, where each cell is a cube (this is probably an optimal shape of the cells). For other boundary shapes, one has two options:

1. Continue using a partitioning of Euclidean space that is independent of the shape of the boundary. This would likely involve enclosing the simulation domain with a cube and then partitioning the cube into cells (some cells would be outside the domain and thus wasted). It is even possible to use the cell method with an infinite simulation domain if hashing techniques are employed [22].

2. Use a cell shape that conforms to the shape of the boundary in some simple way. The shape of the cells will thus change if the boundary deforms during the simulation.

Both options have their pros and cons. It may not be possible to use the second one for very complex boundary shapes. We have used the first approach to generate packings of ellipsoids in a spherical container (useful in comparing with experimental results), by enclosing the spherical container in a cube and partitioning that cube into cells. However, for lattice-based boundaries, we have chosen to use a partitioning of the unit cell into a possibly non-orthogonal Cartesian grid that conforms to the shape of the unit cell. This is illustrated in two dimensions in Fig. 3. The unit cell is partitioned into $N_k^{(c)}$ slabs along each dimension $k = 1, \ldots, d$, to obtain a total of $N_c = \prod_{k=1}^{d} N_k^{(c)}$ identical and consecutively (first along dimension 1, then along dimension 2, etc.) numbered parallelepiped cells. We typically maximize $N_k^{(c)}$ along each dimension such that the distance between the two parallel faces of the cells along any dimension is larger then the extent of the largest particle, $L_c = \min_k L_k > D_{\max}$. Operations in the cell method for lattice-based boundaries basically remain operations on Cartesian grids, just as if the simulation domain had been cubic. Note that each cell has $3^d$ neighbors (including itself), which is only 9 in two, but 27 in three dimensions. As noted earlier, sometimes more then 3 slabs may need to be checked along certain dimensions, depending on the Euclidean cutoff distance for the neighbor search. For completely periodic boundary conditions, there are other schemes for partitioning into cells which preserve the orthogonality and compactness of the unit cell [4], by using alternative choices of the simulation domain. In simulations where the lattice deforms by large amounts, one can alternatively periodically recompute a well-conditioned basis for the lattice and restart the simulation with a new choice of lattice vectors.

*1. The Cell Method in EDMD*

It is useful to briefly sketch the basic ideas of how the cell method is integrated in event-driven algorithms. The cell partitioning is used to speed up the prediction of the next event to happen. This event may be a *boundary event*, which can be a collision of a particle with a hard-wall, or a particle leaving the unit cell in a periodic system. The event may also be a binary collision, and each particle predicts collisions only with the particles in the (first) neighboring cells of its current cell. It is clear that a binary collision cannot occur with a particle not in a neighboring cell until the particle leaves its current cell. Therefore, a boundary event may be a *transfer*, where a particle's centroid leaves its current cell and goes into another cell. The algorithm predicts and processes transfers in time order with the other events.

Whenever a particle undergoes a transfer, it must correct its event prediction. In par-ticular, it must predict binary collisions with all the particles in the *new* neighbor cells. If one maintains separately the prediction for the next boundary *and* the next binary collision, then upon a transfer one can reuse the old binary collision prediction and *only* calculate the collision time with particles in the neighbor cells which were not checked earlier [30]. This cuts the number of neighbor cells to process from $3^d$ to $3^{d-1}$, which can save up to 2/3 in computational effort in three dimensions. In our algorithm we originally maintained the binary collision prediction separately and reused it whenever possible, however, since the neighbor-list method described next is usually superior in practice, we no longer try to reuse previous binary collisions.

## B. The Near-Neighbor List (NNLs) Method

The cell method is the method used in all EDMD algorithms that we are aware of. An exception is the algorithm of Ref. [14], but this algorithm is rather different from the classical EDMD algorithms (and from our algorithm) in more than this respect. There is a preference for the cell method in EDMD because it is very easy to incorporate it into the algorithm, while still maintaining a rigorously provable correct execution of the event sequence, given sufficient numerical precision. For *monodispersed* (equal) spherical particles, particularly at moderate densities, the cell method is truly the best approach. However, for aspherical particles whose aspect ratio is far from 1, the cell method becomes inefficient. This is because one cannot choose the cells small enough to ensure an average of about 1 particle per cell. Instead, due to the large $D_{\max}$, there need to be very few (large) cells which contain many particles and so little computational effort is saved by using the cells. The same is true even for spheres when large polydispersity is present since the cells need to be at least as large as the largest sphere in the system, and therefore there can be many small spheres inside one cell. A more complicated hierarchical cell structure (quadtree or octree) can be used for very polydisperse packings, but such an approach does not directly generalize to nonspherical particles.

In TDMD, a more widely used neighbor search method is the method of *near-neighbor lists* (NNLs) (see for example Ref. [3]). In this method, each particle has a list of its near neighbors, i.e., particles which are in close proximity (for example, within the cutoff for the interaction potential). As the particles move around the lists need to be updated, and this is often done heuristically. Since the particles displace little from time step to time step in TDMD, the lists need to be updated only after many time steps (especially for dense systems). NNLs are not easy to use within EDMD because of the necessity to ensure correctness of the algorithm rigorously. If the order of events is not predicted correctly, the algorithm will typically fail with error conditions such as endless collision cycles between several particles. However, it is easily recognized that in order to efficiently treat nonspherical particles it is necessary to combine neighbor lists with the cell method. We now describe how this can be accomplished while maintaining a provably correct prediction of the collision sequence.

The main drawback of the cell method is that the shape of the cells is not adjusted to the shape of the particles, for example, elongated or squashed particles, but cubic cells. The main advantage, on the other hand, is that the partitioning into cells is static and independent of the motion of the particles. To correct for the drawback, we must compromise on the advantage: The partitioning into "cells" must be updated from time to time to reflect the motion of the particles, if we are to have any hope of having cells which take into account the shape of the particles. The idea is the following: Surround each particle $i$ with a *bounding neighborhood* $\mathcal{N}(i)$, so that the particle is completely inside its bounding neighborhood, and the shape of the neighborhood is in some sense sensitive to the position and shape of the particle (for example, it should be elongated approximately along the same direction as the ellipsoid). Then, consider any two particles whose neighborhoods overlap to be near neighbors, and only calculate interaction potentials or check for collisions between such pairs. Each particle then stores a list of *interactions* in its near-neighbor list $\text{NNL}(i)$, which is equivalent to each bounding neighborhood storing a list of neighborhoods with which it overlaps. This is illustrated for disks by using disks as the bounding neighborhoods in Fig. 4. Note that the cell method, as described earlier, *must* be used when (re-)building the NNLs, since overlap between neighborhoods cannot be checked efficiently otherwise. Building and maintaining the NNLs is expensive and dominates the computation for very aspherical particles. Finally, we note that the choice of the shape of the bounding neighborhoods and the exact way one constructs the NNLs is somewhat of a design choice. The necessary invariant is that each particle be completely contained inside its bounding neighborhood and that there be an interaction in the NNLs for each pair of overlapping neighborhoods.

In this paper we describe a specific conceptually simple approach which applies to hard particles of any shape and has worked well in practice. In our algorithm, the shape of $\mathcal{N}(i)$ is the *same* as the shape of particle $i$, but scaled uniformly with some scaling factor $\mu_{\text{neigh}} > 1$. Additionally, $\mathcal{N}(i)$ has the same centroid as $i$, at least at the instant in time when NNL$(i)$ is constructed (after which the particle may displace). This is illustrated for ellipses in Fig. 4. One wants to have the bounding neighborhood $\mathcal{N}(i)$ as large as possible so that there is more room for the particle $i$ to move without the need to rebuild its NNL. However, the larger the neighborhood, the more neighbors there will be to examine. The optimal balance, as determined by the choice of $\mu_{\text{neigh}}$, is studied numerically in the second part of this series of papers. It is important to note that it would most likely be better to consider $\mathcal{N}(i)$ to be the set of all points that are within a given distance from the surface of particle $i$, especially for very nonspherical particles. This is because scaling a very elongated particle by a given factor $\mu$ produces unnecessarily long neighborhoods, which increases the cost of using the cell method to construct the neighbor lists. However, evaluating point-to-surface or suface-to-surface distances is quite nontrivial even for ellipsoids, and also the geometrical reasoning is obscured. On the other hand, using a bounding neighborhood which has the same shape as the particle is very intuitive and also efficient for ellipsoids, as we show in the second paper in this series.

*1. The NNL Method in EDMD*

Once the NNLs are built, one no longer needs to use the cell method, so long as all particles are still *completely* contained within their bounding neighborhoods. As time progresses, a particle may protrude outside its neighborhood, and in this case the NNLs need to be updated accordingly, using the fail-safe cell method. Details of this update will be given later. Therefore, when using NNLs, instead of transfers, another kind of event needs to be included: a "collision" with its bounding neighborhood. When using NNLs, transfers do not need to be handled at all. Namely, instead of using the cell method for the particles themselves, it should be used on the bounding neighborhoods. Each cell keeps a list of the bounding neighborhoods whose centroids it contains. Hard-walls are handled by including hard walls as neighbors in the NNLs of the particles whose bounding neighborhoods intersect a hard wall. At present we do not try to reuse any previous binary collisions when rebuilding neighbor lists because dealing with such reuse is rather complicated.

An additional complication when using NNLs arises when the boundary is deforming. Since in our approach all positional coordinates are expressed in relation to the (possibly deforming) lattice, the neighborhoods are not stationary but move together with the boundary. This may lead to originally disjoint neighborhoods overlapping later on. In order to ensure correctness of the neighbor search in such cases, one can add a "safety cushion" around each bounding neighborhood $\mathcal{N}(i)$. Specifically, two particles are to be considered neighbors if their bounding neighborhoods overlap when scaled by a common scaling factor $1+\epsilon_N$, where $\epsilon_N > 0$ is the relative size of the safety cushion. The NNLs need to be rebuilt completely whenever the boundary deformation becomes too large, because of the possibility of new neighborhood overlap. In this context, a measure of how much the boundary has deformed is given by the relative amount that Euclidean distances have changed due to the boundary deformation.

Consider a periodic system and two points with relative displacement $\mathbf{r}$, measured in lattice vectors. The Euclidean distance between them is $l^2 = \mathbf{r}^T \mathbf{G} \mathbf{r}$, where $\mathbf{G} = \mathbf{\Lambda}^T \mathbf{\Lambda}$ is a metric tensor. At a later time $\Delta t$, the distance changes, and the largest relative contraction in Euclidean distance between any two points is given by:

$$\min_{\mathbf{r}} \left[ \frac{(l+\Delta l)}{l} \right]^2 = \min_{\mathbf{r}_E} \frac{\mathbf{r}_E^T (\mathbf{I} + \boldsymbol{\epsilon}\Delta t)^T (\mathbf{I} + \boldsymbol{\epsilon}\Delta t) \mathbf{r}_E}{\mathbf{r}_E^T \mathbf{r}_E^T} = \lambda_{\min}\left[(\mathbf{I} + \boldsymbol{\epsilon}\Delta t)^2\right],$$

where $\mathbf{r}_E = \mathbf{\Lambda} \mathbf{r}$ and $\lambda_{\min}$ denotes the minimal eigenvalue of a symmetric matrix. Therefore, the Euclidean distance between the centroids of two neighborhoods would not have contracted by more then a factor of $\lambda_{\min}\left[(\mathbf{I} + \boldsymbol{\epsilon}\Delta t)^2\right]$. In light of this observation, a reasonable heuristic approach is to periodically check the magnitude of the smallest eigenvalue of $(\mathbf{I} + \boldsymbol{\epsilon}\Delta t)^2$, and rebuild the NNLs completely whenever it deviates from unity by more then a few (as determined heuristically via experimentation) multiples of $\epsilon_N$. Since it is reasonable to assume that the boundary deforms slowly compared to the particles, these kinds of updates will happen infrequently. This approach seems to work well in practice. In EDMD a rigorous approach is also possible, by predicting the first instance in time when two non-overlapping bounding neighborhoods first overlap, and including this as a special event in the event queue. When this event is at the top of the queue, the simulation is essentially restarted from the current point in time. However, such an approach does not work in TDMD, and we have not found the practical need for such a complicated scheme either.

## C. Very Aspherical Particles

Using the traditional cell method when rebuilding the NNLs is the computational bottleneck for very aspherical particles, as demonstrated in the second paper of this series. To really obtain a fast yet rigorously correct event-driven algorithm for very aspherical particles the traditional cell method needs to be either abandoned or modified. It is clear that any neighbor search mechanism which only uses the centroids cannot be efficient. Although in a sense Ref. [5] studies the worst case of $\alpha \to \infty$ (needles, and similarly for platelets), it does not mention any additional techniques to handle the fact that as many as 50 needles can be in one cell in the reported simulations. This is probably because at that time only small systems ($N = 100 - 500$) could be studied, for which the cell method does not offer big savings even for spheres.

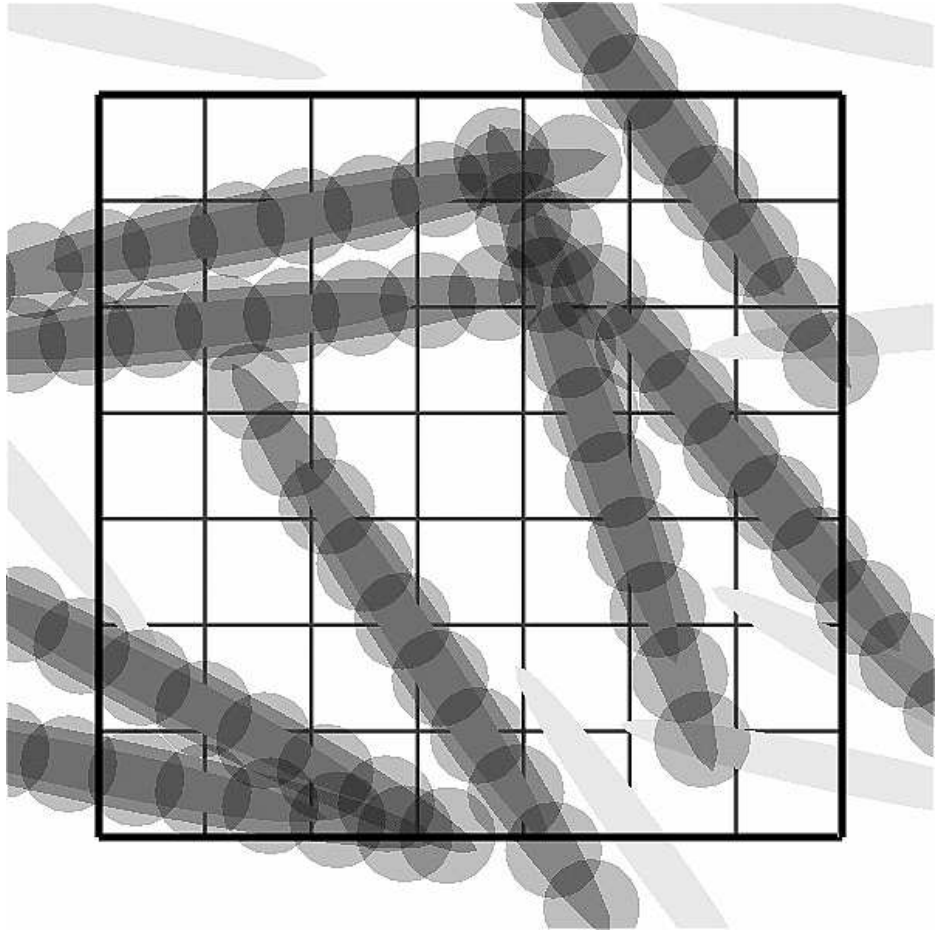

Figure 5: A small periodic packing of ellipses of aspect ratio $\alpha = 10$ illustrating the use of bounding sphere complexes. Each particle $i$ (darkest shade) is bounded by its neighborhood (lighter shade) $\mathcal{N}(i)$, which is itself bounded by a collection of 10 disks BSC($i$). A bounding neighborhood $\mathcal{N}(j)$ may overlap with $\mathcal{N}(i)$ if some of the bounding disks of particles $j$ and $i$ overlap. Therefore the usual cell grid (also shown) can be used in the search for neighbors to add to NNL($i$). Image particles are shown in a lighter shade.

The approach we have implemented is to use several spheres to bound each particle, instead of just one large bounding sphere. We will refer to this collection of bounding spheres as the *bounding sphere complex* (BSC). For the purposes of neighbor search, we still continue to use the cell method, however, we use the cell method on the collection of bounding spheres, not on the particles themselves. That is, we bin all of the bounding spheres in the cells, and the minimal Euclidian length of a cell is at least as large as the largest diameter of a bounding sphere. By increasing the number of bounding spheres per particle one can make the cells smaller. When searching for the neighbors of a given particle, one looks at all of its bounding spheres and their neighboring bounding spheres, and then checks whether the particles themselves are neighbors. This slightly complicates the search for neighbors, but the search can be optimized so that a given pair of particles is only checked once, rather then being checked for every pair of bounding spheres that they may share. It is hard to maintain the binning of the bounding spheres in cells as particles move. It is therefore essential to combine using BSCs with using NNLs. Each bounding neighborhood $\mathcal{N}(i)$ is bounded by BSC($i$), that is, $\mathcal{N}(i)$ is completely contained in the union of the bounding spheres in BSC($i$). The binning of the bounding spheres is only updated when NNL($i$) is updated, and particle $i$ is free to move inside $\mathcal{N}(i)$ without possibility of overlapping with a particle not in NNL($i$). Using BSCs in two dimensions is illustrated in Fig. 5.

In our implementation, we use relative positions and radii for the spheres in BSC($i$), expressed in a coordinate system in which particle $i$'s orientation is aligned with the global coordinate system and the radius of its bounding sphere is unity. This enables us to not have to update the above quantities as the particle moves and changes shape, and also to share them between particles of identical shapes using pointers. When updating $\mathcal{N}(i)$, we can easily calculate the absolute (Euclidian) positions and radii of the bounding spheres from the relative ones.

In two dimensions, for very elongated objects, it is relatively easy to construct bounding complexes, however, this is not so easy in three dimensions, even though there are general methods (taken from computational geometry) for finding a good approximation to a particle shape with a few spheres [13]. We expect that there will be an optimal number of spheres $N_S$ to use, this number increasing as the aspect ratio increases, however, it is not clear how to construct optimal BSCs. The approach we have implemented is to first bound each ellipse or ellipsoid in an orthogonal parallelepiped (rectangle in two dimensions), and then use a subset of a simple cubic lattice cover (a collection of identical spheres whose union covers all of Euclidian space) to bound (cover) the orthogonal parallelepiped. This kind of approach is far from optimal (for example, the lowest density sphere cover in three dimensions is given by a body-centered lattice of spheres), but it is very simple and works relatively well for sufficiently aspherical particles. This is illustrated in three dimensions for prolate and oblate ellipsoids in Fig. 6. As can be seen from the figure, it seems hard, if not impossible, to construct BSCs with few small spheres for flat (oblate) particles. Future research is needed to find a way to speed neighbor search for very oblate particles, and a promising direction to investigate is hierarchical bounding sphere complexes. In the second paper in this series we demonstrate that using BSCs in conjunction with NNLs significantly improve the speed of the EDMD algorithm for very elongated (prolate) particles. Note that using a large number of small bounding spheres (for very aspherical particles) requires a significant increase in the number of cells, and to save memory hashing may need to be used when manipulating the cell partitioning [25].

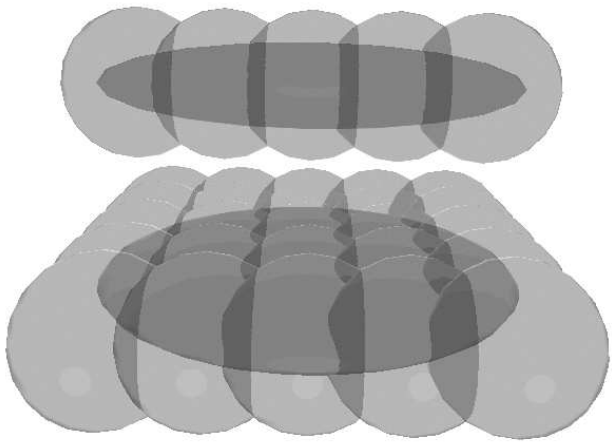

Figure 6: Bounding sphere complexes for spheroids of aspect ratio $\alpha = 5$. The prolate particle has 5 bounding spheres, but the oblate one has 25 bounding spheres.

## IV. EDMD ALGORITHM

In this section, we describe our EDMD algorithm in significant detail, in the hope that this will prove very useful to other researchers implementing similar methods. Starting from a brief history of the main ideas used in the algorithm and a description of the basic notation, we proceed to give detailed descriptions of each step in the algorithm in the form of pseudocodes. We first explain the top level event loop and its most involved step of predicting the impending event for a given particle. We then focus on binary collisions and boundary events separately, and finally describe algorithms for maintaining NNLs in a dynamic environment. Some of the steps of the algorithm, such as predicting the time of collision of two particles or processing a binary collision, depend on the particular particle shape in question and are illustrated specifically for ellipsoids in the second part of this series of papers.

A.  History

We briefly summarize some of the previous work on EDMD algorithms. Although this has been done in other publications, we feel indebted to many authors whose ideas we have used and combined together to produce our algorithm, and would like to acknowledge them.

The very first MD simulation used an event-driven algorithm [1], and since those early attempts the core of an efficient EDMD algorithm for spherical particles, entailing a combination of delayed updates for the particles, the cell method and using a priority queue for the events, has been developed [7, 27]. Our approach borrows heavily from the EDMD algorithm developed by Lubachevsky [19]. We do not use a double-buffering technique as does Lubachevsky, following Ref. [16], and incorporate additional techniques developed by other authors.

One of the controversial questions in the history of EDMD is how many event predictions to retain for each particle $i$? As Ref. [16] demonstrates, it is best to use a heap (complete binary search tree) for the priority event queue, and we follow this approach. It seems clear that only the impending prediction for each particle should be put in the event queue (i.e., the size of the heap is equal to $N$), but this prediction may be invalidated later (due to a third-party event, for example). In such cases, it may be possible to reuse some of the other previously-predicted binary collisions for $i$, for example, the one scheduled with the second-smallest time [23, 29]. This requires additional memory for storing more predictions per particle and adds complexity to the algorithm. We have adapted the conclusion of Ref. [30] that this complexity is not justified from an efficiency standpoint. Ref. [23] makes the important observation that after a transfer fewer cells need to be checked for collisions. The authors of Ref. [30] thus predict and store separately the next binary collision and the next transfer for each particle, and only insert the one with the smaller time into the event heap. More exotic EDMD algorithms, for example, aimed at increased simplicity or ease of vectorization [14], have been developed. We build on these previous developments and combine neighbor-list techniques traditionally used in TDMD to develop a novel EDMD algorithm specifically tailored to systems of nonspherical particles at relatively high densities.

## B. Notation

As explained above, the EDMD algorithm consists of processing a sequence of time-ordered events. Each particle must store some basic information needed to predict and process the events. An *event* $(t_e, p_e)$ is specified by giving the predicted time of occurrence $t_e$ and the partner $p_e$. A special type of an event is a binary collision $[t_c, (p_c, v_c)]$, determined by specifying the time of collision $t_c \equiv t_e$ and the partner in the event $(p_c \equiv p_e, v_c)$. The primary use of the image (virtual) identifier $v_c$ is to distinguish between images of a given particle when periodic boundary conditions are used. Note that the *collision schedules* must be kept symmetric at all times, that is, if particle $i$ has an impending event with $(j, v)$, then particle $j$ must have an impending event with $(i, -v)$. Although the cell a particle belongs to can be determined from the position of its centroid, this is difficult to do exactly when a particle is near the boundary of a cell due to roundoff errors (possible tricks to avoid such problems include adding a cushion around each cell and not considering a transfer until the particle is sufficiently outside the cell [16]). We have chosen to explicitly store and maintain the cell that a particle, a bounding neighborhood of a particle, or a bounding sphere, belongs to (as determined by the corresponding centroid).

In summary, for each particle $i = 1, \ldots, N$, we store:

1. The predicted *impending event* $(t_e, p_e, v_c)$ along with any other information which can help process the event or collision more efficiently should it actually happen later.

2. The last update *time $t$*.

3. The *state* of the particle at time $t$, including:

    (a) Its *configuration*, including the relative *position* of the centroid $\mathbf{r}$ and any additional configuration (such as orientation) $\mathbf{q}$, as well as the particle *shape* (such as radius, semiaxes, etc.) $\mathbf{O}$. Note that $\mathbf{O}$ may be shared among many particles using pointers (for example, all particles have the same shape at all times in a monodisperse packing) and thus not be updated to time $t$ but still be at time zero[4].

---

[4] We have implemented a different approach for systems with a few types of particles (monodisperse, bidisperse, etc.), for which we store the particle shape information separately from the particles and share it among them, and polydisperse systems in which each particle has a (potentially) different shape, for which we store the particle shape together with the rest of the particle state.

(b) The particle *motion*, including the relative velocity of the centroid $\mathbf{v} = \dot{\mathbf{r}}$ and additional (such as angular) velocity $\boldsymbol{\omega}$ representing $\dot{\mathbf{q}}$. Also included in the motion is the rate of deformation of the particle shape $\boldsymbol{\Gamma}$ (possibly shared among different particles).

(c) The particle cell $c$, to which $\mathbf{r}$ belongs, if not using NNLs.

4. Dynamical parameters, such as particle mass or moment of inertia (possible shared with other particles).

5. If using NNLs, the configuration of the (immobile) bounding neighborhood $\mathcal{N}(i)$, $\mathbf{r}_N$ and $\mathbf{q}_N$, its shape $\mathbf{O}_N$, as well as the cell $c$ to which $\mathbf{r}_N$ belongs.

6. If using BSCs in addition to NNLs:

    (a) The relative positions $\mathbf{r}_j^{BS}$ and relative radii $O_j^{BS}$, $j = 1, \ldots, N_{BS}$, of its $N_{BS}$ bounding spheres, along with the largest BS radius $O_{\max}^{BS} = \max_j O_j^{BS}$. These are expressed *relative* to the position and size of $\mathcal{N}(i)$.

    (b) The cell $c_j^{BS}$ that $\mathbf{r}_j^{BS}$ belongs to, $j = 1, \ldots, N_{BS}$.

For each of these quantities, we will usually explicitly indicate the particle to which they pertain, for example, $t(i)$ will denote the time of particle $i$.

1. *Event Identifiers*

Each particle must predict its impending event, and there are several different basic types of events: binary collisions (the primary type of event), wall collisions (i.e., collisions with a boundary of the simulation domain), collisions with a bounding neighborhood (i.e., a particle leaving the interior of its bounding neighborhood), transfers (between cells), and checks (re-predicting the impending event). Additionally, several different types of checks can be distinguished, depending on why a check was required and whether the motion of the particle changed (in which case old predictions are invalid) or not (in which old predictions may be reused). We consider transfers and wall collisions together as *boundary events* (or boundary "collisions"), since their prediction and processing is very similar (especially for periodic BCs). The exact cell wall through which the particle exits the (unit) cell, or the wall with which the particle collides, is identified with an integer $w$, which is negative if the event is with a wall of the unit cell (boundary).

In our implementation, the type of a predicted event for a particle $i$ is distinguished based on the *event partner* $p$ (possibly including an image identifier $v$):

$0 \leq p \leq N$  A binary collision between particles $i$ and $(p, v)$, where $v$ is the virtual identifier of the partner.

$p = -\infty$  A check (update) after an event occured that did not alter the motion of $i$.

$p > 2N$  Transfer between cells, i.e., "collision" with wall $w = p - 2N$, $w > 0$.

$p < -2N$  Wall collision with wall $w = p + 2N$, $w < 0$, which can be a real hard wall or the boundary of the unit cell.

$N < p \leq 2N$  Check after binary collision with partner $(j, -v)$, where $j = p - N$ (the motion of particle $i$ has changed).

$p = 0$  Check for particle $i$ after an event occured which altered the motion of $i$.

$p = \infty$  Collision with the bounding neighborhood $\mathcal{N}(i)$.

The range $-2N \leq p < 0$ is reserved for future (parallel implementation) uses. Of course, one can also store the partner as two integers, one indicating the type of event and the other identifying the partner, however, the above approach saves space.

## C.  Processing the Current Event

Algorithm 1 represents the main event loop in the EDMD algorithm, which processes events one after the other in the order they occur and advances the global time $t$ accordingly. It uses a collection of other auxiliary steps, the algorithms of which are given in what follows. Note that when processing the collision of particle $i$ with particle $(j, v)$, we also update particle $j$, and later, when processing the same collision but as a collision of $j$ and $(i, -v)$, we skip the update. Also, note that when using NNLs, there are two options: Completely rebuild the NNLs as soon as some particle $i$ collides with its neighborhood, or, rebuild only the neighbor list $\text{NNL}(i)$. We discuss the advantages and disadvantages of each approach and compare their practical performance in the second paper in this series.

**Algorithm 1:** Process the next event in the event heap.

1. Delete (pop) the top of the event queue (heap) to find the next particle $i$ to have an event with $p_e(i)$ at $t_e(i)$.

2. Perform global checks to ensure the validity of the event prediction. For example:

    (a) If the boundary is deforming, and if at time $t_e(i)$ the cell length $L_c$ is not larger then the largest enclosing sphere diameter $D_{\max}$, $L_c[t_e(i)] \leq D_{\max}[t_e(i)]$, then restart the simulation:

        i. Synchronize all particles (Algorithm 2).

        ii. Repartition the simulation box to increase the length $L_c$ (for example, for lattice boundaries, increase the appropriate $N_k^{(c)}$).

        iii. Re-bin the particles into the new cells based on the positions of their centroids.

        iv. Reset the event schedule (Algorithm 3).

        v. Go back to step 1.

    (b) If using NNLs and the NNLs are no longer valid (for example, due to boundary deformation), then:

        i. Synchronize all particles.

        ii. Rebuild the NNLs (Algorithm 8).

        iii. Reset the event schedule.

        iv. Go back to step 1.

3. If the boundary is deforming, update its shape. For example, for lattice-based boundaries, set $\mathbf{\Lambda} \leftarrow \mathbf{\Lambda} + \dot{\mathbf{\Lambda}}[t_e(i) - t]$.

4. Advance the global simulation time $t \leftarrow t_e(i)$.

5. If the event to process is not a check after a binary collision, then update the configuration of particle $i$ to time $t$ (for example, $\mathbf{r}(i) \leftarrow \mathbf{r}(i) + [t - t(i)]\mathbf{v}_i$), and set $t(i) \leftarrow t$.

6. If using NNLs and event is a collision with a bounding neighborhood, then:

    (a) If completely rebuilding NNLs, then declare NNLs invalid and execute step 2b.

    (b) Else, record a snapshot of the current shape of particle $i$ (recall that this may be shared with other particles) in $\mathbf{O}_i$ and rebuild the NNL of particle $i$ (Algorithm 9).

7. If the event is a wall collision or cell transfer, then:

    (a) If $p_e(i) > 0$ then set $w \leftarrow p_e(i) - 2N$ (transfer).

    (b) Else set $w \leftarrow p_e(i) + 2N$ (wall collision).

    (c) Process the boundary event with "wall" $w$ (Algorithm 7).

8. If the event is a binary collision, then:

    (a) Update the configuration of particle $j = p_e(i)$ to time $t$ and set $t(j) \leftarrow t$ and $p_e(j) \leftarrow N + i$ (mark $j$'s event as a check).

    (b) Process the binary collision between $i$ and $j$ (see specific algorithm for ellipsoids in second paper in this series).

9. Predict the next collision and event for particle $i$ (Algorithm 4).

10. Insert particle $i$ back into the event heap with key $t_e(i)$.

11. Terminate the simulation or go back to step 1.

Because EDMD is asynchronous, it is often necessary to bring all the particles to the same point in time (synchronize) and obtain a snapshot of the system at the current time $t$. This is done with Algorithm 2. Note that we reset the time to $t = 0$ after such a synchronization step. Another step which appears frequently is to reset all the future event predictions and start afresh, typically after a synchronization. In particular this needs to be done when initializing the algorithm. The steps to do this are outlined in Algorithm 3.

**Algorithm 2** Synchronize all particles to the current simulation time $t$.

1. If $t = 0$ then return.

2. For all particles $i = 1, \ldots, N$ do:

    (a) Update the configuration of particle $i$ to time $t$.

    (b) Set $t_e(i) \leftarrow t_e(i) - t$, $t_c(i) \leftarrow t_c(i) - t$ and $t(i) \leftarrow 0$.

3. Update the shapes of all particles to time $t$.

4. Store the total elapsed time $T \leftarrow T + t$ and set $t \leftarrow 0$.

**Algorithm 3** Reset the schedule of events.

1. Reset the event heap to empty

2. For all particles $i = 1, \ldots, N$ do:

    (a) Set $p_e(i), p_c(i) \leftarrow 0$ and $t_e(i), t_c(i) \leftarrow 0$.

    (b) Insert particle $i$ into the event heap with key $t_e(i)$.

## D. Predicting The Next Event

The most important and most involved step in the event loop is predicting the next event to happen to a given particle, possibly right after another event has been processed. Algorithm 4 outlines this process. Note that it is likely possible to further extend and improve this particular step by better separating motion-altering from motion-preserving events and improving the reuse of previous event predictions.

**Algorithm 4:** Predict the next binary collision and event for particle $i$, after an event involving $i$ happened.

1. If not using NNLs, then:

    (a) Initialize $t_w \leftarrow \infty$ and $\tilde{t}_w \leftarrow \infty$ and set $w \leftarrow 0$.

    (b) Predict the next boundary event (wall collision or transfer) time $t_w$ and partner "wall" $w$ for particle $i$, if any, by looking at all of the boundaries of $c(i)$ (Algorithm 6). If an exact prediction could not be made (for example, if a hard wall was involved and the search was terminated prematurely), calculate a time $\tilde{t}_w$ up to which a boundary event is guaranteed not to happen and set $w \leftarrow 0$.

    (c) If $w = 0$, then force a check at time $\tilde{t}_w$, $p_e(i) \leftarrow -\infty$ and $t_e(i) \leftarrow \tilde{t}_w$,

    (d) else predict $t_e(i) \leftarrow t_w$ and:

        i. If $w < 0$ then set $p_e(i) \leftarrow w - 2N$,

        ii. else set $p_e(i) \leftarrow w + 2N$.

    (e) If a hard-wall prediction was made, store any necessary information needed to process the collision more efficiently later (for example, store $\lambda$ in the case of ellipsoids, as explained in the second paper in this series).

    (f) For all particles $(j, v)$ in the cells in the first neighborhood of $c(i)$, execute step 4,

2. else if using NNLs, then:

    (a) Predict the time $t_N$ particle $i$ will protrude outside of (collide with) its bounding neighborhood $\mathcal{N}(i)$, limiting the length of the search interval to $t_e(i)$. If an exact prediction is not possible, calculate a time $\tilde{t}_N$ before which $i$ is completely contained in $\mathcal{N}(i)$.

    (b) If $t_N$ was calculated and $t_N < t_e(i)$, then record:

        i. Set $p_e(i) \leftarrow \infty$ and $t_e(i) \leftarrow t_N$.

        ii. Potentially store any additional information about this collision for particle $i$,

    (c) else if $\tilde{t}_N$ was calculated and $\tilde{t}_N < t_e(i)$ then force a new prediction for particle $i$ at time $\tilde{t}_N$, $p_e(i) \leftarrow -\infty$ and $t_e(i) \leftarrow \tilde{t}_N$.

    (d) For all hard walls $w$ in NNL$(i)$, predict the time of collision $t_w$. If an exact prediction could not be made, calculate a time $\tilde{t}_w$ up to which the collision is guaranteed not to happen.

    (e) If $t_w$ was calculated and $t_w < t_e(i)$, then record:

        i. Set $t_e(i) \leftarrow t_w$ and $p_e(i) \leftarrow w - 2N$.

        ii. Potentially store any necessary information needed to process the wall collision more efficiently later,

    (f) else if $\tilde{t}_w$ was calculated and $\tilde{t}_w < t_e(i)$, then force a check $p_e(i) \leftarrow -\infty$ and $t_e(i) \leftarrow \tilde{t}_w$.

    (g) For all particles $(j, v)$ in NNL$(i)$, execute step 4,

3. Skip step 4.

4. Predict the time of collision between particles $i$ and $(j, v)$:

    (a) Predict if $i$ and $(j, v)$ will collide during a time interval of length $\min[t_e(i), t_e(j)]$ and if yes, calculate the time of collision $t_c$, or calculate a time $\tilde{t}_c < t_c$ before which a collision will not happen (see specific algorithm for ellipsoids in the second paper in this series).

    (b) If $t_c$ was calculated and $t_c < \min[t_e(i), t_e(j)]$, then record this collision as the next predicted binary collision for particle $i$:

        i. Set $p_e(i) \leftarrow j$, $v(i) \leftarrow v$ and $t_e(i) \leftarrow t_c$.

        ii. Potentially store any additional information about this collision for particle $i$ (for example, $\lambda$ in the case of ellipsoids),

- (c) else if $\tilde{t}_c$ was calculated and $\tilde{t}_c < t_e(i)$ then force a new prediction for particle $i$ at time $\tilde{t}_c$, $p_e(i) \leftarrow -\infty$ and $t_e(i) \leftarrow \tilde{t}_c$.

5. If $0 < p_e(i) \leq N$ then let $j = p_e(i)$ (a new collision partner was found), and:

   (a) If the involved third-party $m = p_e(j)$ is a real particle, $0 < m \leq N$ and $m \neq i$, then invalidate the third party collision prediction, $p_e(m) \leftarrow -\infty$.

   (b) Ensure that the collision predictions are symmetric by setting $p_e(j) \leftarrow i$, $v(j) = -v(i)$ and $t_e(j) \leftarrow t_e(i)$. Also copy any additional information about the predicted collision to particle $j$ as well (in the case of ellipsoids, this involves storing $(1 - \lambda)$ for particle $j$).

   (c) Update the key of $j$ in the event heap to $t_e(j)$.

---

### E. Binary Collisions

The two main steps in dealing with binary collisions is predicting them and processing them. Processing a collision is inherently tied to the shape of the particle. We give a generic specification of how to predict binary collisions between particles in Algorithm 5, and a specific implementation for ellipsoids is given in the second part of this series of papers.

### F. Boundary Events

In this section we focus on lattice-based boundaries and give a prescription for predicting and processing boundary events (transfers and wall collisions).

#### 1. Prediction

When NNLs are not used, one must check all the boundaries of the current particle cell $c(i)$ and find the first time the particle leaves the cell or collides with a hard wall, if any. We do not give details for predicting or processing hard-wall collisions in this paper. For lattice based boundaries, the prediction of the next boundary event proceeds independently along each dimension, and then the smallest of the $d$ event times is selected, as illustrated in Algorithm 6.

**Algorithm 5** Predict the (first) time of collision between particles $i$ and $(j, v)$, $t_c$. If prediction cannot be verified, return a time $\tilde{t}_c$ before which a collision will not happen. Possibly also return additional information about the collision.

1. Convert $v$ into a virtual displacement of particle $j$ in terms of unit cells, $\Delta \mathbf{r}_j \equiv \mathbf{n}_c$, as discussed in Section II C.

2. Calculate the current configuration of the particles $i$ and $j$, for example, the positions of their centroids,

$$\mathbf{r}_i \leftarrow \mathbf{r}(i) + [t - t(i)] \mathbf{v}_i$$
$$\mathbf{r}_j \leftarrow \mathbf{r}(j) + [t - t(j)] \mathbf{v}_j + \Delta \mathbf{r}_j,$$

and their current orientations for nonspherical particles.

3. If the shape of the particles is changing, calculate the current shape of $i$ and $j$.

4. Eliminate any further use of relative positions in the procedure by calculating the current Euclidean positions, velocities and accelerations of the particles using Eqs. (4-6) and the above $\mathbf{r}_i$ and $\mathbf{r}_j$.

5. Calculate the collision time $t_c$ or $\tilde{t}_c$ of two moving and possibly deforming particles of the given initial shapes and configurations and initial Euclidean positions, velocities and accelerations, assuming a force-free motion starting at time zero. Optionally collect additional information needed to process the collision faster if it actually happens. See the specific algorithm for ellipsoids in the second paper in this series.

6. Correct the prediction to account for the current time, $t_c \leftarrow t_c + t$ or $\tilde{t}_c \leftarrow \tilde{t}_c + t$.

*2. Processing*

Processing the boundary events amounts to little work when the event is a transfer from one cell to another. For periodic BCs however, additional work occurs when the particle crosses the boundary of the unit cell (i.e., the simulation domain), since in this case it must be translated by a lattice vector in order to return it back into the unit cell. Considerably more complicated is the processing of collisions with hard walls, especially for nonspherical particles or when the lattice velocity is nonzero, however, we do not give the details of these steps in Algorithm 7.

**Algorithm 6** Predict the next *wall event* with "partner" $w$ for particle $i$ moving with relative velocity $\mathbf{v}(i)$ and the time of occurrence $t_w$, for a *lattice-based boundary*. The sign of $w$ determines the type of event: $w > 0$ specifies that particle $i$ leaves its cell $c(i)$ through one of the cell boundaries, while $w < 0$ specifies that the particle collides with one of the hard walls or crosses one of the boundaries of a unit cell *and* leaves its bin, for a periodic system. The value of $|w|$ determines the exact cell boundary or wall.

1. Convert the cell identifier $1 \leq c(i) \leq N_c$ into a $d$-dimensional vector giving the positions of the cell in the Cartesian grid of cells, $1 \leq \mathbf{g}^{(c)} \leq \mathbf{N}^{(c)}$.

2. For all dimensions, $k = 1, \ldots, d$, do:

    (a) Predict the time when the particle centroid will cross a wall of $c(i)$ along dimension $k$:

    i. If $v_k(i) = [\mathbf{v}(i)]_k > 0$ (particle will exit on the "right" side of the bin), then

    A. Set $w_k \leftarrow 2(k-1) + 2$ and
    $$t_k \leftarrow \left[\mathbf{g}_k^{(c)} - N_k^{(c)} r_k(i)\right] / \left[N_k^{(c)} v_k(i)\right].$$

    B. If boundary is periodic along dimension $k$ and $\mathbf{g}_k^{(c)} = N_k^{(c)}$, then set $w_k \leftarrow -w_k$,

    ii. else if $v_k(i) < 0$ (particle will exit on the "left" side of the bin), then

    A. Set $w_k \leftarrow 2(k-1) + 1$, and
    $$t_k = \left[N_k^{(c)} r_k(i) - \mathbf{g}_k^{(c)} + 1\right] / \left[N_k^{(c)} v_k(i)\right].$$

    B. If boundary is periodic along dimension $k$ and $\mathbf{g}_k^{(c)} = 1$, then set $w_k \leftarrow -w_k$,

    iii. else set $t_k \leftarrow \infty$ and $w_k \leftarrow 0$.

    (b) If boundary is not periodic along dimension $k$, then also predict the time of collision with the hard wall boundaries along dimension $k$, assuming that the particle starts from zero time:

    i. If $\mathbf{g}_k^{(c)} = N_k^{(c)}$, predict time of collision with the "right" hard wall along dimension $k$, $t_k^{(hw)}$. If $t_k^{(hw)} < t_k$, then set $t_k \leftarrow t_k^{(hw)}$ and $w_k \leftarrow -[2(k-1) + 2]$.

    ii. If $\mathbf{g}_k^{(c)} = 1$, predict time of collision with the "left" hard wall along dimension $k$, $t_k^{(hw)}$. If $t_k^{(hw)} < t_k$, then set $t_k \leftarrow t_k^{(hw)}$ and $w_k \leftarrow -[2(k-1) + 1]$.

3. Find the dimension $\tilde{k}$ with the smallest $t_k$ and return $t_w = t(i) + t_{\tilde{k}}$ and $w = w_{\tilde{k}}$.

**Algorithm 7** Process the boundary event (transfer or collision with a hard-wall) of particle $i$ with wall $w$, assuming a lattice-based boundary.

1. From $w$, find the dimension $k$ along which the event happens and the side ("left" or "right").

2. If this is a boundary event, $w < 0$, then:

   (a) If the boundary is periodic along $k$, then:

      i. Shift the particle by a unit cell, $r_k(i) \leftarrow r_k(i) + 1$ if particle is exiting its cell to the right, or $r_k(i) \leftarrow r_k(i) - 1$ if exiting to the left.

      ii. Let $j = p_c(i)$. If $0 < j \leq N$, correct the virtual identifiers for the predicted collision between $i$ and $j$, $v(i)$ and $v(j)$, to account for the shift in step 2(a)i.

      iii. Pretend that this is a simple transfer, $w \leftarrow 2N - w$,

   (b) else, process the collision of the particle with the hard-wall. This will typically involve calculating the Euclidean position and velocity of the particle, calculating the exchange of momentum between the particle and the wall, calculating the new Euclidean velocity of the particle $\mathbf{v}^{(E)}$ (and also $\boldsymbol{\omega}$ if necessary), converting back to relative velocity, and updating the velocity $\mathbf{v}$ (and $\boldsymbol{\omega}$).

3. If this is a transfer (note step 2(a)iii above), $w > 0$, then update the cell of the particle $c(i)$ and move the particle from the linked list of its previous cell to the list of the new cell.

## G. Building and Updating the NNLs

In our implementation, all of the NNLs are implemented as an optimized form of linked lists. Each interaction $[(j, v), p]$ in NNL$(i)$ stores the *partner* $(j, v)$ and a *priority* $p$. We usually prescribe a fixed upper bound on the number of neighbors (interactions) $N_i$ that a particle can have (this allows us to preallocate all storage and guarantee that additional memory will not be used unless really necessary), which can vary between particles if necessary. Only the $N_i$ interactions with highest priority are retained in NNL$(i)$. This kind of NNL can be used for a variety of tasks, including finding the first few nearest neighbors of any particle. We allow the NNLs to asymmetric, i.e., just because particle $i$ interacts with particle $(j, k)$, it is not implied that particle $j$ interacts with $(i, -k)$, but rather, the reverse interaction must be stored in NNL$(j)$ if needed. In the particular use of NNLs for neighbor search, the priorities are the negative of the "distances" between the particles, so that only the closest $N_i$ particles are retained as neighbors.

There are two main ways of updating the NNLs after a particle collides with its bounding neighborhood. One is to completely update the NNLs of all particles and start afresh, and the other one is to only update the NNL of the particle in question. We next discuss these two forms of NNL updates, *complete* and *partial*, and compare them practically in the second paper in this series to conclude that it is in general preferable to use partial updates (however, there are situations when it is best to use complete updates). As explained earlier, we focus on the case when the bounding neighborhoods are scaled versions of the particles. In addition to limiting the number of near-neighbors of any particle to $N_i$, we limit the maximum scaling of the neighborhood with respect to the particle itself to $\mu_{\text{cutoff}} \geq \mu_{\text{neigh}} > 1$, and count as overlapping any neighborhoods which overlap when scaled by an additional factor $(1 + \epsilon_\mu)$, where $\epsilon_\mu \geq 0$ is a safety cushion used when the boundary deforms. Henceforth, denote $\mu_{\text{max}} = (1 + \epsilon_\mu)\,\mu_{\text{cutoff}}$.

*1. Complete Updates*

A simpler form of update is after a complete resetting of the NNLs, i.e., building the NNLs from scratch. Algorithm 8 gives a recipe for this. The aim of the algorithm is to try to make the bounding neighborhoods have a scale factor of $\mu_{\text{max}}$ and add all overlapping neighborhoods in the NNLs. This will always be possible if $N_i$ is large enough. However, we allow one to limit the number of near neighbors. This is useful when there is not a good estimate of what a good $\mu_{\text{cutoff}}$ is.

The algorithm is significantly more complicated when BSCs are used since the search for possibly overlapping bounding neighborhoods needs to be done over pairs of bounding spheres. To avoid checking a given pair of bounding neighborhoods for overlap multiple times, we use an integer mask $M(i)$ for each particle, which we assume is persistent, i.e., stored for each particle between updates. In our algorithm, a hard wall can be a neighbor in NNL($i$) if $\mathcal{N}(i)$ is intersected by a hard-wall boundary. For simplicity, we do not present pseudocode for adding these hard-wall neighbors, however, it is a straightforward exercise to add these steps to the algorithms below.

**Algorithm 8:** *Completely* update the near-neighbor lists (NNLs) by rebuilding them from scratch. Assume all particles have been synchronized to the same point in time.

1. For all particles, $i = 1, \ldots, N$, reset $\text{NNL}(i)$ to an empty list.

2. For all particles, $i = 1, \ldots, N$, reduce $\mathbf{r}(i)$ to the first unit cell, and if $\mathbf{r}(i)$ is no longer inside $c(i)$, then remove $i$ from the linked list of $c(i)$, update $c(i)$, and insert $i$ in the list of the new $c(i)$.

3. If using BSCs, then initialize the largest (absolute) radius of a bounding sphere $O^{(E)}_{\max} \leftarrow 0$, and for all particles, $i = 1, \ldots, N$, do:

    (a) Set the bounding neighborhood of $i$ to have the same centroid, orientation and shape as $i$ but be scaled by a factor $\mu_{\max}$, $\mathbf{r}_N(i) \leftarrow \mathbf{r}(i)$, $\mathbf{q}_N(i) \leftarrow \mathbf{q}(i)$ and $\mathbf{O}_N(i) \leftarrow \mu_{\max} \mathbf{O}(i)$.

    (b) For all bounding spheres of $i$, $k = 1, \ldots, N_{BS}(i)$, do:

        i. Remove the sphere from the linked list of cell $c^{BS}_k(i)$.

        ii. Calculate the new absolute position of its center and the cell it is in, update $c^{BS}_k(i)$ accordingly, and insert the sphere into the linked list of $c^{BS}_k(i)$.

        iii. Calculate the absolute radius $O^{(E)}_k$ of the bounding sphere and set $O^{(E)}_{\max} \leftarrow \max\left\{O^{(E)}_{\max}, O^{(E)}_k\right\}$.

    (c) Initialize the mask $M(i) \leftarrow 0$.

4. else let $O^{(E)}_{\max} \leftarrow \mu_{\max} \{\max_i [O_{\max}(i)]\}$ be the largest possible radius of an enclosing sphere of a bounding neighborhood.

5. For all particles, $i = 1, \ldots, N$, do:

    (a) If using BSCs, then for all bounding spheres of $i$, $k = 1, \ldots, N_{BS}(i)$, do:

        i. For all cells $c_i$ in the neighborhood of $c^{BS}_k(i)$ of Euclidean extent $2O^{(E)}_{\max}$, and for all bounding spheres in $c_i$ belonging to some particle $(j, v)$, do:

            A. If $j \geq i$ and $M(j) \neq \text{sign}(v)(|v|N + i)$, then execute step 7,

            B. else mark this pair of particles as already checked, $M(j) \leftarrow \text{sign}(v)(|v|N + i)$.

    (b) else, for all cells $c_i$ in the neighborhood of $c(i)$ of Euclidean extent $2O^{(E)}_{\max}$ (note that his may involve higher-order neighbors of $c(i)$), do:

        i. For all particles $(j, v) \in c_i$ such that $j \geq i$, execute step 7.

6. Skip step 7.

7. If the largest common scaling factor which leaves $i$ and $(j, v)$ disjoint, $\mu_{ij} \leq \mu_{\max}$, then:

    (a) Calculate $\mu_{ij}$ exactly (ellipsoids are treated in the second paper in this series).

        i. Insert the interaction $[(j, v), -\mu_{ij}]$ in NNL$(i)$. Note this may remove some previous entries in NNL$(i)$ if it is already full.

        ii. Insert the interaction $[(i, -v), -\mu_{ij}]$ in NNL$(j)$.

8. For all particles, $i = 1, \ldots, N$, do:

    (a) Initialize the minimal scaling of $i$ which makes it overlap with the bounding neighborhood of a non-neighbor particle, $\mu_{\min}^{\text{non-neigh}} \leftarrow \mu_{\max}$.

    (b) If NNL$(i)$ is not full, then initialize the maximal scaling of $i$ which leaves it disjoint from at least one of the bounding neighborhoods of a neighbor particle, $\mu_{\max}^{\text{neigh}} \leftarrow \mu_{\max}$, otherwise initialize $\mu_{\max}^{\text{neigh}} \leftarrow 0$.

    (c) For all interactions $[(j, v), p]$ in NNL$(i)$, ensure that they are bi-directional:

        i. If $-p > \mu_{\max}^{\text{neigh}}$, then set $\mu_{\max}^{\text{neigh}} \leftarrow -p$.

        ii. If there is no interaction with particle $(i, -v)$ in NNL$(j)$, then:

            A. If $-p > \mu_{\min}^{\text{non-neigh}}$, then set $\mu_{\min}^{\text{non-neigh}} \leftarrow -p$.

            B. Delete the interaction $[(j, v), p]$ from NNL$(i)$.

        iii. If $-p < \mu_{\text{neigh}}$, then set $\mu_{\text{neigh}} \leftarrow -p$.

    (d) Set $\mu_{\text{neigh}} \leftarrow \min\left(\mu_{\min}^{\text{non-neigh}}, \mu_{\max}^{\text{neigh}}\right)$. Note that if NNL$(i)$ never filled up then $\mu_{\text{neigh}} = \mu_{\max}$.

    (e) If $\mu_{\text{neigh}} < \mu_{\max}$, then set $\mathbf{O}_N(i) \leftarrow \mu_{\max} \mathbf{O}(i)$.

    (f) If using BSCs and $\mu_{\text{neigh}} < \mu_{\max}$, then for all bounding spheres of $i$, $k = 1, \ldots, N_{BS}(i)$, do:

        i. Remove the sphere from the linked list of cell $c_k^{BS}(i)$.

        ii. Calculate the new absolute position of its center and the cell it is in, update $c_k^{BS}(i)$ accordingly, and insert the sphere into the linked list of $c_k^{BS}(i)$.

9. If the boundary is deforming, record the current shape of the boundary to be used later to verify the validity of the NNLs (see Section III B 1).

## 2. Partial Updates

A considerably more complex task is updating NNL($i$) while trying to leave the lists of other particles intact, other then possibly adding or deleting an interaction involving $i$. We give a prescription for this in Algorithm 9, but do not give many details, as understanding each step is not necessary to get an idea of the overall approach. For simplicity, we do not present the case when BSCs are used, as the modifications to allow for bounding complexes closely parallel those in Algorithm 8 and it is a straightforward exercise for the reader to modify the algorithm below accordingly.

**Algorithm 9:** Update the near-neighbor list of particle $i$, NNL($i$). Assume that the current shape of $i$ is passed in $\mathbf{O}_i$.

1. For all interactions with $(j, v)$ in NNL($i$), delete the reverse interaction with $(i, -v)$ in NNL($j$).

2. Initialize the minimal scaling of $i$ which makes it overlap with the bounding neighborhood of a non-neighbor particle, $\mu_{\min}^{\text{non-neigh}} \leftarrow \mu_{\max}$, as well as the maximal scaling of $i$ which leaves it disjoint from at least one of the bounding neighborhoods of a neighbor particle, $\mu_{\max}^{\text{neigh}} \leftarrow \mu_{\max}$.

3. For all cells $c_i$ in the neighborhood of $c(i)$ of Euclidean extent $O_{\max}^{\text{neigh}} + \mu_{\max} O_i$, where $O_i$ is the radius of the bounding sphere of $i$ and $O_{\max}^{\text{neigh}}$ is the radius of the largest enclosing sphere of a particle neighborhood, do:

    (a) For all bounding neighborhoods $\mathcal{N}_j$ in the list of $c_i$, $c(j) = c_i$, do:

    i. If the largest scaling factor which leaves $i$ disjoint from the neighborhood of $(j, v)$, $\mu_{ij}^{\text{neigh}} < \mu_{\max}$, then calculate $\mu_{ij}^{\text{neigh}}$ exactly, else continue with next particle $(j, v)$.

    ii. If there is room in NNL($j$), then insert the interaction $\left[(j, v), -\mu_{ij}^{\text{neigh}}\right]$ in NNL($i$),

    iii. else if $\mu_{ij}^{\text{neigh}} < \mu_{\min}^{\text{non-neigh}}$ then set $\mu_{\min}^{\text{non-neigh}} \leftarrow \mu_{ij}^{\text{neigh}}$.

    iv. If $\mu_{ij}^{\text{neigh}} < \mu_{\max}^{\text{neigh}}$ then set $\mu_{\max}^{\text{neigh}} \leftarrow \mu_{ij}^{\text{neigh}}$.

4. For all interactions $[(j, v), p]$ in NNL($i$), do:

    (a) Insert the interaction $[(i, -v), -p]$ in NNL($j$).

    (b) If $-p > \mu_{\max}^{\text{neigh}}$ then set $\mu_{\max}^{\text{neigh}} \leftarrow -p$.

5. If NNL($i$) is full, then set $\mu_{\text{neigh}} \leftarrow \mu_{\text{min}}^{\text{non-neigh}}$,

6. else set $\mu_{\text{neigh}} \leftarrow \min\left(\mu_{\text{min}}^{\text{non-neigh}}, \mu_{\text{max}}^{\text{neigh}}\right)$.

7. Set $\mathbf{r}_N(i) \leftarrow \mathbf{r}(i)$, $\mathbf{q}_N(i) \leftarrow \mathbf{q}(i)$ and $\mathbf{O}_N(i) \leftarrow \mu_{\text{neigh}}\mathbf{O}(i)$, and also update $O_{\text{max}}^{\text{neigh}} \leftarrow \max\left[O_{\text{max}}^{\text{neigh}}, \mu_{\text{neigh}} O_i\right]$.

## V. CONCLUSION

In this first paper in a series of two papers, we presented a serial collision-driven molecular dynamics algorithm for nonspherical particles, with a specific focus on improving the efficiency by developing novel techniques for neighbor search. In particular, we developed a rigorous scheme that incorporates near-neighbor lists into event-driven algorithms, and further improved the handling of very elongated objects via the use of (non-hierarchical) bounding sphere complexes. We gave detailed pseudocodes to illustrate the major steps of the algorithm. All necessary details to implement the algorithm for ellipses and ellipsoids are given in the second paper in this series, along with a discussion of the practical performance of the algorithm. Acknowledgments are given in the second paper in this series.

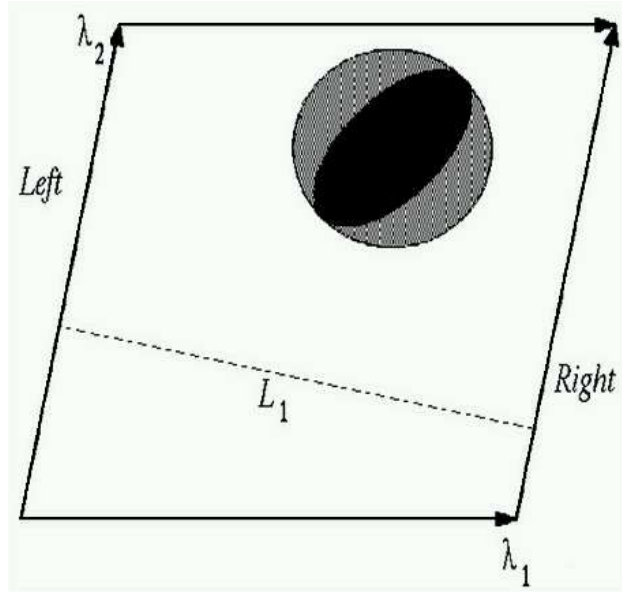

(a) Lattice-based boundary

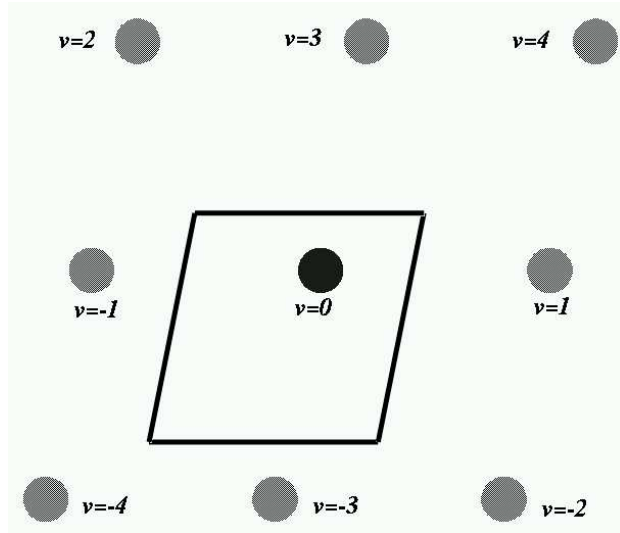

(b) Periodic BCs

Figure 1: Illustration of lattice-based boundaries. The top subfigure shows a unit cell in two dimensions, along with the length of the unit cell along the $x$ direction, $L_1$, and the left and right "walls" along the $x$ dimension. Also shown is a particle and its bounding sphere (disk). The bottom subfigure shows a unit cell and an original particle (black), along with the first neighbor images and their image identifiers $v$.

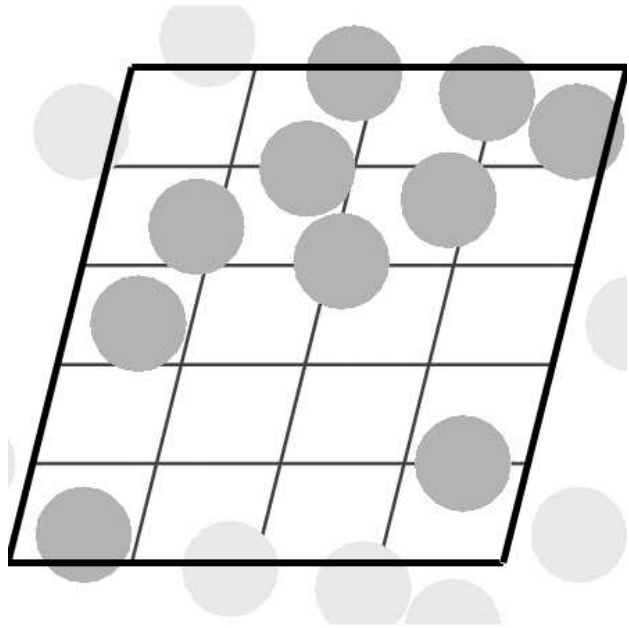

Figure 2: *The cell method*: A small disk packing and the associated grid of cells, to be used in searching for possibly overlapping particles. Dark-shaded disks are original particles and light-shaded ones are images.

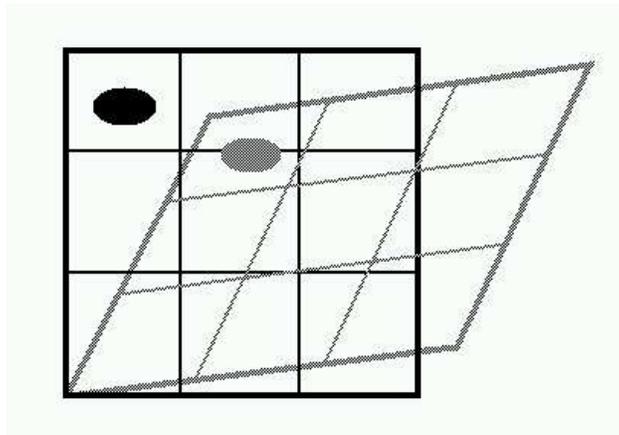

Figure 3: The partitioning of a lattice-based simulation box into cells (dark lines). The Cartesian grid of cells deforms in unison with the lattice, as illustrated by a snapshot of the box and its partitioning at a latter time. Particles also move together with the lattice, even if they are at "rest", $\mathbf{v} = 0$, as shown for an ellipsoid in the $(1,3)$ cell.

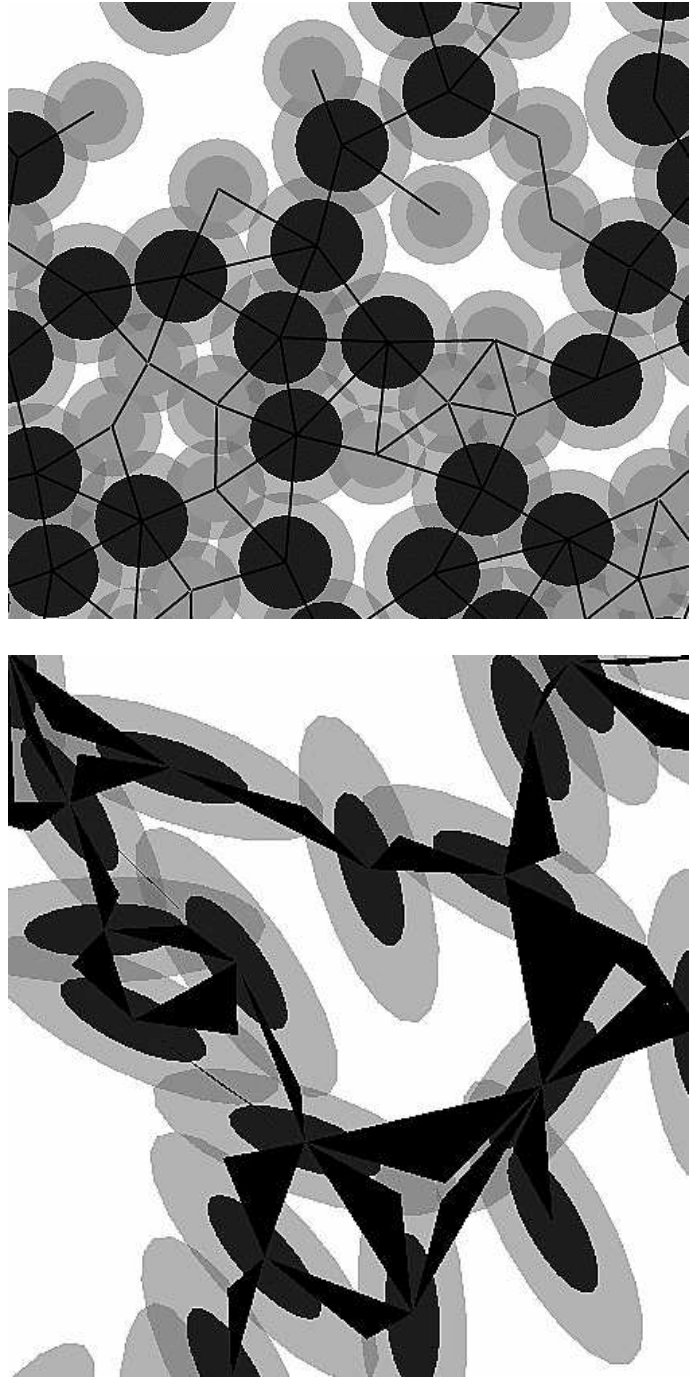

Figure 4: Illustration of NNLs for a system of disks (top) and ellipses (bottom). Particles are dark, and their bounding neighborhoods are light (it is easy to see which neighborhood goes with which particle). For disks, the neighborhoods are disks and the pairs of near-neighbors are shown as dark lines. For ellipses the neighborhoods are ellipses themselves and the interactions are shown as dark triangles whose vertices are given by the centroids of the two ellipses and the point of contact of the ellipses.

# Neighbor List Collision-Driven Molecular Dynamics Simulation for Nonspherical Hard Particles.
# II. Applications to Ellipses and Ellipsoids

Aleksandar Donev,[1,2] Salvatore Torquato,[1,2,3,*] and Frank H. Stillinger[3]

[1]*Program in Applied and Computational Mathematics,*
*Princeton University, Princeton NJ 08544*
[2]*Materials Institute, Princeton University, Princeton NJ 08544*
[3]*Department of Chemistry, Princeton University, Princeton NJ 08544*

## Abstract

We apply the algorithm presented in the first part of this series of papers to systems of hard ellipses and ellipsoids. The theoretical machinery needed to treat such particles, including the overlap potentials, is developed in full detail. We describe an algorithm for predicting the time of collision for two moving ellipses or ellipsoids. We present performance results for our implementation of the algorithm, demonstrating that for dense systems of very aspherical ellipsoids the novel techniques of using neighbor lists and bounding sphere complexes, offer as much as two orders of magnitude improvement in efficiency over direct adaptations of traditional event-driven molecular dynamics algorithms. The practical utility of the algorithm is demonstrated by presenting several interesting physical applications, including the generation of jammed packings inside spherical containers, the study of contact force chains in jammed packings, and melting the densest-known equilibrium crystals of prolate spheroids.

---

[*] Electronic address: `torquato@electron.princeton.edu`

## I. INTRODUCTION

In the first paper of this series of two papers, we presented a collision-driven molecular dynamics algorithm for simulating systems of nonspherical hard particles. The algorithm rigorously incorporates near-neighbor lists, and further improves the treatment of very elongated objects via the use of bounding sphere complexes. Detailed pseudocodes for the algorithm were presented, but several particle-shape-dependent components were left unspecified. In particular, a key component of the algorithm is the evaluation of overlap between scaled versions of two particles, such as the evaluation of the minimal common scaling that leaves two disjoint ellipsoids nonoverlapping, or the maximal scaling of an ellipsoid which leaves it contained within another ellipsoid. Additionally, the required procedures for predicting the time-of-collision for two moving ellipsoids, as well as processing the collision, are developed. Moreover, we discuss generalizations to other particle shapes.

We also illustrate the practical utility and versatility of the algorithm by presenting several nontrivial and physically relevant applications. In particular, we show that by incorporating particle growth (i.e., shape deformation), the proposed algorithm can generate *jammed packings* of ellipsoids and is superior to previously used algorithms in both speed and particularly in accuracy. The high precision of the event-driven approach enables us to reach previously unavailable high densities and to produce tightly jammed ellipsoid packings with several thousand particles, even for relatively large aspect ratios [12]. Additionally, the inclusion of boundary deformation allowed us to generate the densest known crystal packings of ellipsoids [11], and here we show how the algorithm can be used to simulate a quasi-equilibrium (adiabatic) cooling of this crystal to track its equation of state and phase behavior. We also demonstrate how using near-neighbor lists can help monitor the particle collision history near the jamming point and enable the study of *force chains*, previously studied only in time-driven molecular dynamics of soft particles [3].

We present the tools necessary to rapidly evaluate overlap functions for ellipsoids in Section II. We then describe the missing particle-shape-dependent pieces of the algorithm in Section III. Some performance results for the algorithm are shown in Section IV, particularly focusing on the use of our near-neighbor list and bounding sphere complexes techniques. Three illustrative applications are given in Section V.

## II. GEOMETRY OF ELLIPSES AND ELLIPSOIDS

In this section, we focus exclusively on ellipsoidal particles, particularly in two (ellipses) and three (ellipsoids) dimensions. We present all of the necessary tools to adapt the EDMD algorithm to ellipsoids. We first give some introductory material, and then discuss several overlap (contact) functions for ellipsoids, based on the work of Perram and Wertheim [30]. We then focus on calculation of these overlap functions and their time derivatives, which is used in Section III to robustly determine the time-of-collision for two moving ellipsoids. We will attempt to present most of the results so that they generalize to other dimensions as well; however, this is not always possible. We present the basic concepts in a unified and simple manner. Readers looking for more detailed background information are referred to Refs. [1] and [2].

### A. Introduction and Background

In order to deal with the rotational degrees of freedom for ellipsoids and track their orientation, as well as their centroidal position, some additional machinery is necessary. Since very little of the notation seems to be standard, we first present our own notational system, which attempts to unify different dimensionalities whenever possible. We then discuss orientational degrees of freedom and rotation of rigid bodies.

#### 1. Notation

When dealing with rotational motions, especially in three dimensions, cross products and rotational matrices appear frequently. In order to unify our presentation for two and three dimensions as much as possible, we introduce some special matrix notation. Matrix multiplication is assumed whenever products of matrices or a matrix and a vector appear. We prefer to use matrix notation whenever possible[1] and do not carefully try to distinguish between scalars and matrices of one element. We denote the dot product $\mathbf{a} \cdot \mathbf{b}$ with $\mathbf{a}^T\mathbf{b}$, and the outer product $\mathbf{a} \otimes \mathbf{b}$ with $\mathbf{ab}^T$.

---

[1] Our computational implementation uses a specially designed library of inlined macros for matrix operations extensively.

The first notational difficulty relates to the notion of a cross product. In three dimensions, there is only the familiar cross product

$$\mathbf{a} \times \mathbf{b} = \begin{bmatrix} a_y b_z - a_z b_y \\ a_z b_x - a_x b_z \\ a_x b_y - a_y b_x \end{bmatrix} = \mathbf{A}\mathbf{b} \qquad (1)$$

where

$$\mathbf{A} = \begin{bmatrix} 0 & -a_z & a_y \\ a_z & 0 & -a_x \\ -a_y & a_x & 0 \end{bmatrix} = -\mathbf{A}^T$$

is a skew-symmetric matrix which is characteristic of the cross product and is derived from a vector. We will simply *capitalize* the letter of a vector to denote the corresponding *cross product matrix* (like $\mathbf{A}$ above corresponding to $\mathbf{a}$). In two dimensions however, there are two "cross products". The first one gives the velocity of a point $\mathbf{r}$ in a system which rotates around the origin with an angular frequency $\boldsymbol{\omega}$ (which has just one $z$ component and can also be considered a scalar $\omega$),

$$\mathbf{v} = \boldsymbol{\omega} \boxtimes \mathbf{r} = \begin{bmatrix} -\omega \mathbf{r}_y \\ \omega \mathbf{r}_x \end{bmatrix} = \boldsymbol{\Omega}\mathbf{r}, \qquad (2)$$

where

$$\boldsymbol{\Omega} = \begin{bmatrix} 0 & -\omega \\ \omega & 0 \end{bmatrix} = -\boldsymbol{\Omega}^T$$

is a cross product matrix derived from $\boldsymbol{\omega}$. The second kind of "cross product" gives the torque around the origin of a force $\mathbf{f}$ acting at a point (arm) $\mathbf{r}$,

$$\boldsymbol{\tau} = \mathbf{f} \times \mathbf{r} = -\mathbf{r} \times \mathbf{f} = [f_x r_y - f_y r_x] = \mathbf{F}^L \mathbf{r}, \qquad (3)$$

where

$$\mathbf{F}^L = \begin{bmatrix} -f_y & f_x \end{bmatrix} = -\left(\mathbf{F}^R\right)^T$$

is another cross product matrix derived from a vector (the $L$ and $R$ stand for left and right multiplication, respectively). Note that in three dimensions all of these coincide, $\mathbf{F}^L = \mathbf{F}^R = \mathbf{F}$, and also $\boxtimes \equiv \times$. The notation was chosen so equations look simple in three dimensions, but are also applicable to two dimensions. The wedge product generalizes the cross product in higher dimensions [20].

## 2. Rigid Bodies

Representing the orientation of a rigid body in a computationally convenient way has been a subject of debate in the past [2]. A rigid body has

$$f_R = \frac{d(d-1)}{2} = \begin{cases} 1 \text{ if } d = 2 \\ 3 \text{ if } d = 3 \end{cases} \quad (4)$$

rotational degrees of freedom, and this is the minimal number of coordinates needed to specify the configuration of a hard nonspherical particle, in addition to the usual $d$ coordinates needed to specify the position of the centroid. In two dimensions orientations are easy to represent via the angle $\phi$ between the major semiaxes of the ellipsoid and the $x$ axis. But in three dimensions specifying three (Euler) angles is numerically unstable, and extensive experience has determined that for MD computationally the best way to represent orientations is via *normalized quaternions*, which in fact represent finite *rotations* starting from an initial reference configuration (but see Ref. [17] for a discussion). In the case of ellipsoids this reference configuration is one in which all semiaxes are aligned with the coordinate axes. In two dimensions we use a normalized complex number to represent orientation, but for simplicity we will sometimes use the term "quaternion" in both two and three dimensions. Higher dimensional generalizations are discussed in Ref. [33].

In three dimensions, normalized quaternions consist of a scalar $s$ and a vector $\mathbf{p}$,

$$\mathbf{q} = [s, \mathbf{p}] = \left[\cos\frac{\phi}{2}, \left(\sin\frac{\phi}{2}\right)\hat{\boldsymbol{\phi}}\right], \quad (5)$$

where $\hat{\boldsymbol{\phi}}$ is the unit vector along the axis of rotation and $\phi$ is the angle of rotation around this axis, and the normalization condition

$$\|\mathbf{q}\|^2 = s^2 + \|\mathbf{p}\|^2 = 1$$

is satisfied. Therefore in three dimensions we use 4 numbers to represent orientation, which seems like wasting one floating-point number. It is in fact possible to represent the rotation with the oriented angle $\boldsymbol{\phi} = \phi\hat{\boldsymbol{\phi}}$, which is just a vector with 3 coordinates. However, such a representation has numerical problems when $\phi = 0$, and also the representation is not unique[2]. More importantly, combining rotations (as during rotational motion) does not correspond (as one may expect) to vector addition of the $\phi$'s, but it does correspond to quaternion multiplication of the $\mathbf{q}$'s, which is fast since there is no need of repeating the trigonometric evaluations. This is the reason why we also use quaternions in two-dimensions, and represent the orientation of a particle in the plane with 2 coordinates (components of a unit complex number),

$$\mathbf{q} = [s, p] = [\cos \phi, \sin \phi]. \tag{6}$$

---
[2] Note that the quaternion representation is also not unique since $-\mathbf{q}$ and $\mathbf{q}$ represent the same orientation.

The orthogonal *rotation matrix* corresponding to the rotation described by the quaternion (5) is given with

$$\mathbf{Q} = 2 \left[ \mathbf{p}\mathbf{p}^T - s\mathbf{P} + \left(s^2 - \frac{1}{2}\right) \mathbf{I} \right]$$

in three dimensions, and

$$\mathbf{Q} = \begin{bmatrix} s & p \\ -p & s \end{bmatrix}$$

in two dimensions, corresponding to the complex number (6). The resulting orientation after first the rotation $\mathbf{Q}_1$ is applied and then the rotation $\mathbf{Q}_2$ is applied, $\mathbf{Q}_{12} = \mathbf{Q}_2\mathbf{Q}_1$, is represented by the quaternion product

$$\mathbf{q}_{12} = \mathbf{q}_1\mathbf{q}_2 = [s_1 s_2 - \mathbf{p}_1 \cdot \mathbf{p}_2, s_1\mathbf{p}_2 + s_2\mathbf{p}_1 - \mathbf{p}_1 \times \mathbf{p}_2] \tag{7}$$

in three dimensions, and by the complex number product

$$\mathbf{q}_{12} = \mathbf{q}_1\mathbf{q}_2 = [s_1 s_2 - p_1 p_2, s_1 p_2 + s_2 p_1] \tag{8}$$

in two dimensions.

In this work we are interested in particles which move continuously in time. The rate of rotation of a rigid body is given by the *angular velocity* $\boldsymbol{\omega}$ (which can also be considered a scalar $\omega$ in two dimensions), or equivalently, the infinitesimal change in orientation is given by the infinitesimal rotation $d\boldsymbol{\phi} = \boldsymbol{\omega}dt$. The instantaneous time derivative of the normalized quaternion is given with[3]

$$\dot{\mathbf{q}} = \frac{1}{2} \begin{bmatrix} s & -\mathbf{p} \\ \mathbf{p} & s\mathbf{I} + \mathbf{P} \end{bmatrix} \begin{bmatrix} 0 \\ \boldsymbol{\omega} \end{bmatrix}$$

in three dimensions, and with

$$\dot{\mathbf{q}} = \frac{1}{2} \begin{bmatrix} s & p \\ -p & s \end{bmatrix} \begin{bmatrix} 0 \\ \omega \end{bmatrix}$$

---

[3] Recall that capital $\mathbf{P}$ denotes the cross product matrix corresponding to $\mathbf{p}$.

in two dimensions. The time derivative of the corresponding rotation matrix is

$$\dot{\mathbf{Q}} = -\mathbf{Q}\mathbf{\Omega},$$

and this result has was used extensively in deriving the various time derivatives related to the contact function for ellipsoids, as will be given shortly.

## 3. Ellipsoids

An ellipsoid is a smooth convex body consisting of all points $\mathbf{r}$ that satisfy the quadratic inequality

$$(\mathbf{r} - \mathbf{r}_0)^T \mathbf{X} (\mathbf{r} - \mathbf{r}_0) \leq 1, \tag{9}$$

where $\mathbf{r}_0$ is the position of the center (centroid), and $\mathbf{X}$ is a characteristic *ellipsoid matrix* describing the shape and orientation of the ellipsoid. The case when $\mathbf{X} = \frac{1}{O^2}\mathbf{I}$ is a diagonal matrix describes a sphere of radius[4] $O$, which does not require orientation information. In the general case,

$$\mathbf{X} = \mathbf{Q}^T \mathbf{O}^{-2} \mathbf{Q}, \tag{10}$$

where $\mathbf{Q}$ is the rotational matrix describing the orientation of the ellipsoid, and $\mathbf{O}$ is a diagonal matrix containing the major semi-axes of the ellipsoid along the diagonal. The time derivative of the matrix (10) for an ellipsoid rotating with instantaneous angular velocity $\boldsymbol{\omega}$ is

$$\dot{\mathbf{X}} = \mathbf{\Omega}\mathbf{X} - \mathbf{X}\mathbf{\Omega}. \tag{11}$$

In Algorithm 1 we give a prescription for updating the orientation of an ellipsoid rotating with a *constant* angular velocity for a time $\Delta t$.

---
**Algorithm 1** Update the orientation of an ellipsoid rotating with a uniform angular velocity $\boldsymbol{\omega}$ after a time step $\Delta t$.

---
1. Calculate the change in orientation $\mathbf{q}_{\Delta t}$ using $\phi_{\Delta t} = \boldsymbol{\omega}\Delta t$ in eq. (5) or (6).

2. Update the quaternion, $\mathbf{q} \leftarrow \mathbf{q}\mathbf{q}_{\Delta t}$, using eq. (7) or (8).

---

[4] We will use the letters $r$ and $R$ to denote positions of points, and therefore resort to using $O$ when referring to radius.

3. If $|\|\mathbf{q}\| - 1| > \epsilon_{\mathbf{q}}$ (due to accumulation of numerical errors), renormalize the quaternion,

$$\mathbf{q} \leftarrow \mathbf{q}/\|\mathbf{q}\|.$$

## B. Ellipsoid Overlap Potentials

The problem of determining whether two ellipsoids $A$ and $B$ overlap (have a common point) or not has been considered previously in relation to Monte Carlo or MD simulations of hard-ellipsoid systems [1]. Here we are concerned not only with a binary overlap criterion, but rather with a numerically efficient way of measuring a *distance*[5] between the two ellipsoids $F(A, B)$, whose sign not only gives us an overlap criterion,

$$\begin{cases} F(A, B) > 0 \text{ if } A \text{ and } B \text{ are disjoint} \\ F(A, B) = 0 \text{ if } A \text{ and } B \text{ are externally tangent} \\ F(A, B) < 0 \text{ if } A \text{ and } B \text{ are overlapping,} \end{cases}$$

but which is also continuously differentiable in the positions and orientations of the ellipsoids $A$ and $B$ and is numerically stable. An additional convenient property is that $F(A, B)$ be defined and easy to compute for *all* positions and orientations of the ellipsoids. We will call such a distance function an *overlap potential*. We will also make use of an overlap potential $G(A, B)$ for the case when ellipsoid $A$ is completely contained within $B$ (for example, $B$ can be the bounding neighborhood of $A$, or it can be an ellipsoidal hard-wall container),

$$\begin{cases} G(A, B) > 0 \text{ if } A \text{ is completely contained in } B \\ G(A, B) = 0 \text{ if } A \text{ is internally tangent to } B \\ G(A, B) < 0 \text{ if part or all of } A \text{ is outside } B, \end{cases}$$

and give such a potential below. Such potentials have not been considered before since they do not appear in other algorithms, however, our neighbor-list EDMD algorithm for ellipsoids uses it to construct bounding neighborhoods for the particles, and additionally, such a potential can be used to implement hard-wall boundary conditions inside an ellipsoidal container.

---

[5] This is not a distance in the mathematical sense.

More than three decades ago, Vieillard-Baron proposed an overlap criterion based on the number of negative eigenvalues of a certain matrix [44], and this criterion has been subsequently rediscovered [45]. It easily generalizes to two dimensions and can be used to obtain an overlap potential. We have implemented and tested this overlap potential but have found it both computationally and theoretically inferior to an overlap potential proposed by Perram and Wertheim [30]. We have therefore completely adapted the Perram-Wertheim (PW) overlap potential and also extended it to the case of one ellipsoid contained within another. Many other approaches are possible, for example, an approximate measure of the Euclidean distance between the surfaces of the two ellipsoids can be used [14, 16, 23, 31]. However, the advantage of the PW approach is its inherent symmetry, dimensionless character, and most of all, its simple geometric interpretation in terms of scaling factors.

The geometrical idea behind the Perram-Wertheim overlap potential is very simple and is based on considering scaling the size of the ellipsoids uniformly until they are in external or internal tangency. Consider for example the case when $A$ and $B$ are disjoint, as illustrated in the leftmost part of Fig. 1. If ellipsoid $A$ is scaled by a nonnegative factor $\mu(A)$ such that the centroid of $B$ is still outside it, then there is a corresponding scaling of $B$, $\mu(B)$, which brings $B$ into external tangency with $A$ at the *contact point* $\mathbf{r}_C[\mu(A)]$. This scaling is a solution to a simple eigenvalue-like problem involving $\mathbf{X}_A$ and $\mathbf{X}_B$. The normal vectors of $A$ and $B$ at the contact point are of opposite direction, and by changing the ratio of their lengths from $0$ to $\infty$ we get a path of contact points going from the center of $A$ to the center of $B$. It was a wonderful idea of Perram and Wertheim [30] to parameterize this path with a scalar $\lambda \in [0, 1]$, and then look for the $\lambda = \Lambda$ which makes $\mu(A) = \mu(B)$, i.e. look for the common scaling factor $\mu_{AB}$ which brings $A$ and $B$ into external tangency at the contact point $\mathbf{r}_C$ (shown in Fig. 1), or equivalently, look for the largest common scaling factor which preserves non-overlap. This approach is very well-suited for the case when both $A$ and $B$ are particles and thus should be treated equally. Sometimes, however, ellipsoid $B$ has a special status, for example, it may be the bounding neighborhood of another particle. In this case we look for the scaling factor $\mu_B(A)$ of $A$ which brings $A$ into external tangency with the fixed $B$ (see second subsection in Section II.C in Ref. [29]), or equivalently, the largest scaling of $A$ which preserves non-overlap, as illustrated in the middle part of Fig. 1. A similar idea applies to the case when $A$ is contained within $B$, in which case we look for the largest scaling $\nu_B(A)$ of $A$ which leaves $A$ contained completely within $B$, or equivalently, which brings $A$ into internal tangency with $B$.

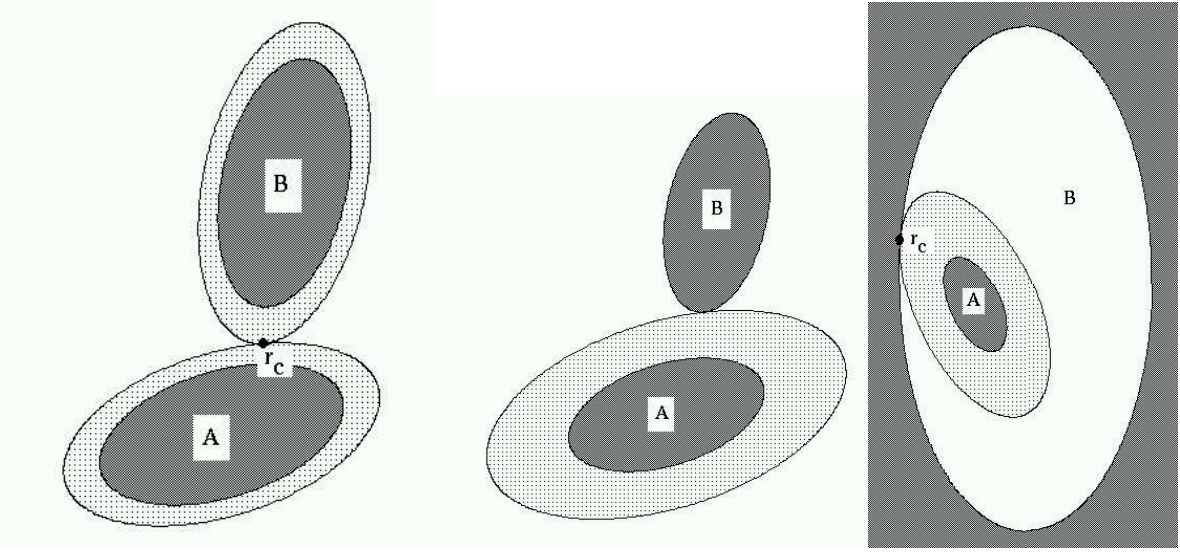

Figure 1: Illustration of the scaling $\mu$ in the PW contact function: *Left*: The outer tangency potential $\mu_{AB}$. *Middle*: The outer tangency potential $\mu_B(A)$. *Right:* The inner tangency potential $\nu_B(A)$.

Using these scaling factors, we can define several overlap potentials,

$$F_{AB}(A,B) = \mu_{AB}^2 - 1 \qquad (12)$$

$$F_B(A,B) = \mu_B^2(A) - 1 \qquad (13)$$

$$G_B(A,B) = \nu_B^2(A) - 1, \qquad (14)$$

which we will refer to as the Perram-Wertheim (PW), the modified PW overlap potential, and the internal PW overlap potential respectively. To appreciate why we use the squares of the scaling factors, consider the case of spheres, where the PW overlap potential simply becomes

$$F_{AB} = \frac{|\mathbf{r}_A - \mathbf{r}_B|^2}{(O_A + O_B)^2} - 1 = \frac{l_{AB}^2}{(O_A + O_B)^2} - 1,$$

which avoids the use of square roots in calculating the distance between the centers of $A$ and $B$, $l_{AB}$, and is also much simpler to work with analytically. Extensive use of all three of the contact functions (12-14) has been made in the implementation of the algorithm, and in particular, the building and updating of the near-neighbor lists. The original PW overlap potential (12) is the most efficient in practice and also has the property that it is symmetric with respect to the interchange of $A$ and $B$, and is preferred over (13) unless $\mu_B^2(A)$ is needed (recall from the first paper in this series that $\mu_B(A)$ is needed when using partial updates for the neighbor lists). Note that $F_B$ and $G_B$ are not defined for all positions of the ellipsoids, namely, if the center of $A$ is inside $B$, $F_B$ is not defined, and conversely, if the center of $A$ is outside $B$, $G_B$ is not defined.

## C.  Calculating the Overlap Potentials

In this section we address the issue of efficiently and reliably calculating the three PW overlap potentials. We base our discussion on outlines of recipes for calculating $F_{AB}$ and $F_B$ in the literature [1, 29, 30], but focus on detail and describe a specific computational scheme based on polynomials. Additionally, contact information such as the point of contact or the common normal vector at the point of contact can be calculated once the overlap potential is found.

### 1.  Evaluating $F_{AB}$

Following Perram and Wertheim, define the parametric function

$$f_{AB}(\lambda) = \lambda(1-\lambda)\mathbf{r}_{AB}^T\mathbf{Y}^{-1}\mathbf{r}_{AB}, \tag{15}$$

where $\mathbf{r}_{AB} = \mathbf{r}_B - \mathbf{r}_A$, and

$$\mathbf{Y} = \lambda\mathbf{X}_B^{-1} + (1-\lambda)\mathbf{X}_A^{-1}. \tag{16}$$

It turns out that this function is strictly concave on the interval $[0, 1]$ and thus has a unique maximum at $\lambda = \Lambda \in [0, 1]$, from which one can directly calculate the overlap potential:

$$F_{AB} = f_{AB}(\Lambda) = \max_{0\leq\lambda\leq 1} f_{AB}(\lambda).$$

The maximum of $f_{AB}(\lambda)$ can easily be found numerically using only polynomial manipulations, by making extensive use of matrix adjoints (sometimes called adjugates) and determinants (both of which are polynomials in the matrix elements). First rewrite $f_{AB}(\lambda)$ as a rational function:

$$f_{AB}(\lambda) = \frac{p_{AB}(\lambda)}{q_{AB}(\lambda)} = \frac{\lambda(1-\lambda)\left\{\mathbf{a}_{AB}^T\text{adj}\left[\lambda\mathbf{I} + (1-\lambda)\mathbf{A}_{AB}\right]\mathbf{a}_{AB}\right\}}{\det\left[\lambda\mathbf{I} + (1-\lambda)\mathbf{A}_{AB}\right]}, \tag{17}$$

where
$$\mathbf{a}_{AB} = \mathbf{X}_B^{1/2}\mathbf{r}_{AB} \text{ and } \mathbf{A}_{AB} = \mathbf{X}_B^{1/2}\mathbf{X}_A^{-1}\mathbf{X}_B^{1/2}.$$

Note that powers of $\mathbf{X}$ are easy to calculate because of the special form (10) and orthogonality of $\mathbf{Q}$. We have made use of the symbolic algebra system Maple® and its code generation abilities to generate inlined Fortran code to form the coefficients of the polynomial $\text{adj}\left[\lambda\mathbf{I} + (1-\lambda)\mathbf{A}\right]$ and $\det\left[\lambda\mathbf{I} + (1-\lambda)\mathbf{A}\right]$ for a given symmetric matrix $\mathbf{A}$, and this has found numerous uses when dealing with ellipsoids, such as in evaluating the coefficients of the polynomials $p_{AB}$ and $q_{AB}$ in eq. (17). The unique maximum of $f_{AB}(\lambda)$ can be found by finding the root of its first derivative, which is the same as finding the unique root of the degree-$2d$ polynomial

$$h_{AB} = p'_{AB}q_{AB} - p_{AB}q'_{AB}$$

in the interval $[0, 1]$, which can be done very rapidly using a safeguarded Newton method.

A good initial guess to use in Newton's method is the exact result for spheres

$$\Lambda = \frac{\overline{O}_A}{\overline{O}_A + \overline{O}_B},$$

where $\overline{O}$ is the largest semiaxes, i.e., the radius of the enclosing sphere for an ellipsoid. Additionally, one often has a better initial guess for $\Lambda$ in cases when the relative configuration of the ellipsoids has not changed much from previous evaluations of $F_{AB}$. Finally, a task which appears frequently is to evaluate the overlap potential between two ellipsoids but only if they are closer than a given cutoff, in the sense that the exact value is only needed if $F_{AB} \leq F_{AB}^{(\text{cutoff})}$, or equivalently $\mu_{AB} \leq \mu_{AB}^{(\text{cutoff})}$ (see for example the algorithms for updating the neighbor lists in the first paper in this series). This cutoff can be used to speed up the process by terminating the search for $\Lambda$ as soon as a value $f_{AB}(\lambda) > F_{AB}^{(\text{cutoff})}$ is encountered during Newton's method. Additionally, one can first test the enclosing spheres for $A$ and $B$ with the same cutoff and not continue the calculation if the spheres are disjoint even when scaled by a factor $\mu_{AB}^{(\text{cutoff})}$.

Since almost always the value of $\lambda = \Lambda$ is used, henceforth we do not explicitly denote the special value $\Lambda$, unless there is the possibility for confusion. The reader should keep in mind that expressions to follow are to be evaluated at $\lambda = \Lambda$. The subscript $C$ will be used to denote quantities pertaining to the contact point. The *contact point* $\mathbf{r}_C$ of the two ellipsoids is

$$\mathbf{r}_C = \mathbf{r}_A + (1-\lambda)\mathbf{X}_A^{-1}\mathbf{n} = \mathbf{r}_B - \lambda\mathbf{X}_B^{-1}\mathbf{n}, \tag{18}$$

where

$$\mathbf{n} = \mathbf{Y}^{-1}\mathbf{r}_{AB} \qquad (19)$$

is the unnormalized common *normal vector* at the point of contact (once the ellipsoids are scaled by the common factor $\mu_{AB}$), directed from $A$ to $B$ in this case. Here $\mathbf{r}_{BC} = \mathbf{r}_C - \mathbf{r}_B$ and $\mathbf{r}_{AC} = \mathbf{r}_C - \mathbf{r}_A$ are the "arms" from the centers of the ellipsoids to the contact point. An important value is the curvature of $f_{AB}$ at the special point $\lambda = \Lambda$,

$$f_{\lambda\lambda} = \frac{d^2 f_{AB}}{d\lambda^2} = 2\frac{\mathbf{r}_{BC}^T \mathbf{Y}^{-1}\mathbf{r}_{AC}}{\lambda(1-\lambda)} = -2\mathbf{n}^T \mathbf{Z}\mathbf{n} < 0,$$

where

$$\mathbf{Z} = \mathbf{X}_A^{-1}\mathbf{Y}^{-1}\mathbf{X}_B^{-1} = \mathbf{X}_B^{-1}\mathbf{Y}^{-1}\mathbf{X}_A^{-1} = [\lambda \mathbf{X}_A + (1-\lambda)\mathbf{X}_B]^{-1}.$$

2. *Evaluating $F_B$ and $G_B$*

The evaluation of the modified outer and internal tangency PW overlap potentials $F_B$ and $G_B$ proceeds in a similar fashion, but with a differing sign in several expressions. Here the upper sign will denote the case of internal tangency ($G_B$), and the lower the case of outer tangency ($F_B$). We proceed to give a prescription for evaluation of these potentials without detailed explanations.

As for evaluating $F_{AB}$ above, first we define the parameterized function

$$f_B(\lambda) = \lambda^2 \mathbf{r}_{AB}^T \mathbf{Y}^{-1} \mathbf{X}_B^{-1} \mathbf{Y}^{-1} \mathbf{r}_{AB}, \qquad (20)$$

as well as

$$g_B(\lambda) = (1-\lambda)^2 \mathbf{r}_{AB}^T \mathbf{Y}^{-1} \mathbf{X}_A^{-1} \mathbf{Y}^{-1} \mathbf{r}_{AB}, \qquad (21)$$

where

$$\mathbf{Y} = \lambda \mathbf{X}_B^{-1} \mp (1-\lambda)\mathbf{X}_A^{-1}. \qquad (22)$$

We then numerically look for the largest $\lambda = \Lambda$ in $[0,1]$ which solves the nonlinear equation

$$f_B(\lambda) = 1, \qquad (23)$$

and then we have the desired scaling factor

$$G_B \text{ or } F_B = g_B(\Lambda) = \nu_B^2 - 1 \text{ or } \mu_B^2 - 1.$$

Additionally, the contact point is

$$\mathbf{r}_C = \mathbf{r}_A \mp (1-\lambda)\mathbf{X}_A^{-1}\mathbf{n} = \mathbf{r}_B \pm \lambda\mathbf{X}_B^{-1}\mathbf{n}, \qquad (24)$$

where the normal vector $\mathbf{n}$ is as in eq. (19). An additional useful value is the slope

$$f_\lambda = \frac{df_B}{d\lambda} = 2\frac{\mathbf{r}_{BC}^T \mathbf{Y}^{-1} \mathbf{r}_{AC}}{(1-\lambda)}.$$

We can again use polynomial algebra to efficiently solve eq. (23) using a safeguarded Newton method, by rewriting $f_B(\lambda)$ as a rational function

$$f_B(\lambda) = \frac{\sum_{k=1}^d \left[\lambda p_k^{(B)}(\lambda)\right]^2}{q_B^2(\lambda)} = \frac{\lambda^2 \left\|\operatorname{adj}\left[\lambda \mathbf{I} \mp (1-\lambda)\mathbf{A}_{AB}\right]\mathbf{a}_{AB}\right\|^2}{\det\left[\lambda \mathbf{I} \mp (1-\lambda)\mathbf{A}_{AB}\right]} = 1, \qquad (25)$$

where $p_k^{(B)}$ and $q_B$ are polynomials, to obtain the equivalent equation

$$\frac{q_B(\lambda)}{\lambda\sqrt{\sum_{k=1}^d \left[p_k^{(B)}(\lambda)\right]^2}} = 1,$$

which is better suited for numerical solution. The search interval for $\Lambda$ in this case should be taken to be $\left[\tilde{\Lambda}, 1\right]$, where $\tilde{\Lambda}$ is the largest root of the degree-$d$ polynomial $q_B$ in $[0, 1]$, which can be found exactly in both two and three dimensions using standard algebraic methods for the solution of polynomial equations of degree less than 5. A reasonable initial guess when evaluating $G_B$ is $\lambda = \tilde{\Lambda}$.

Unlike the evaluation of $F_{AB}$ and $F_B$, which are both rapid[6] and robust, the evaluation of $G_B$ poses numerical difficulties due to the presence of the minus sign in eq. (22), which can cause $\mathbf{Y}$ to become singular. This happens when $\|\mathbf{Y}^{-1}\mathbf{r}_{AB}\| \to 0$, which does occur when $\Lambda \to \tilde{\Lambda}$. In this case, since $\mathbf{Y}$ is singular, its adjoint is (almost always) rank-1,

$$\operatorname{adj}[\mathbf{Y}] \to \mathbf{u}\mathbf{u}^T,$$

where $\mathbf{u}$ is some (eigen)vector, and the problem occurs because $\mathbf{u}^T \mathbf{r}_{AB} \to 0$, yielding an apparently indeterminate $0/0$ in eq. (25). The limiting value of $G_B$ is mathematically well-defined even in this case, however, its numerical evaluation is unstable, and has been a constant source of numerical problems in our implementation. One alleviating trick is to avoid explicitly inverting $\mathbf{Y}$ and instead the adjoint should be used, $\mathbf{Y}^{-1} = \text{adj}\left[\mathbf{Y}\right]/\det\left[\mathbf{Y}\right]$, where the determinant of $\mathbf{Y}$ can be calculated by using (23),

$$\det\left[\mathbf{Y}\right] = \lambda\sqrt{\tilde{\mathbf{n}}^T \mathbf{X}_B^{-1} \tilde{\mathbf{n}}},$$

where $\tilde{\mathbf{n}} = \text{adj}\left[\mathbf{Y}\right]\mathbf{r}_{AB}$. Even with such precautions, we have observed numerical difficulties in the calculations involving inner tangency of $A$ and $B$. It would therefore be useful to explore alternative overlap potentials for the case when ellipsoid $A$ is contained within ellipsoid $B$, or different ways of calculating $G_B$.

---

[6] In our numerical experience $F_{AB}$ and its time derivatives can be evaluated significantly faster.

## D. Time Derivatives of the Overlap Potentials

When dealing with moving ellipsoids, and in particular, when determining the time-of-collision for two ellipsoids in motion, expressions for the time derivatives of the contact potentials are needed. We give these expressions here without a detailed derivation. We have additionally obtained expressions for second order derivatives, however, these are not needed for the current exposition and are significantly more complicated, and are not presented here. We use the standard dot notation for time derivatives.

### 1. Derivatives of $F_{AB}$

Consider two ellipsoids moving with instantaneous velocities $\mathbf{v}_A$ and $\mathbf{v}_B$ and rotating with instantaneous angular velocities $\boldsymbol{\omega}_A$ and $\boldsymbol{\omega}_B$. For the purposes of the Lubachevsky-Stillinger algorithm, we also want to allow the ellipsoid semiaxes to change with an expansion/contraction rate of $\boldsymbol{\gamma} = \dot{\mathbf{O}}$, i.e., $\mathbf{O}(t) = \mathbf{O}(0) + \boldsymbol{\gamma}t$. We have the expected result that the rate of change of overlap depends on the *projection* of the relative velocity at the point of contact, $\mathbf{v}_C$, along the common normal vector $\mathbf{n}$,

$$\dot{F}_{AB} = 2\lambda\left(1 - \lambda\right)\mathbf{n}^T \mathbf{v}_C, \tag{26}$$

where

$$\mathbf{v}_C = [\mathbf{v}_B + \boldsymbol{\omega}_B \boxtimes \mathbf{r}_{BC} + \boldsymbol{\Gamma}_B \mathbf{r}_{BC}] - [\mathbf{v}_A + \boldsymbol{\omega}_A \boxtimes \mathbf{r}_{AC} + \boldsymbol{\Gamma}_A \mathbf{r}_{AC}]$$

and

$$\boldsymbol{\Gamma} = \mathbf{Q}^T \left(\mathbf{O}^{-1}\boldsymbol{\gamma}\right)\mathbf{Q}.$$

One sometimes also needs the time derivative of $\lambda = \Lambda$

$$\dot{\lambda} = -\frac{2}{f_{\lambda\lambda}} \left\{ \tilde{\mathbf{n}}^T \mathbf{v}_C + \mathbf{n}^T \left[ \lambda \mathbf{\Gamma}_B \mathbf{r}_{BC} + (1-\lambda) \mathbf{\Gamma}_A \mathbf{r}_{AC} \right] + \lambda (1-\lambda) \mathbf{n}^T \mathbf{Z} \left[ (\boldsymbol{\omega}_B - \boldsymbol{\omega}_A) \boxtimes \mathbf{n} \right] \right\}, \tag{27}$$

where

$$\tilde{\mathbf{n}} = \lambda \mathbf{Y}^{-1} \mathbf{r}_{BC} + (1-\lambda) \mathbf{Y}^{-1} \mathbf{r}_{AC}.$$

*2. Derivatives of $F_B$ and $G_B$*

In this case we have:

$$\dot{G}_B \text{ or } \dot{F}_B = \mp 2(1-\lambda) \mathbf{n}^T \mathbf{v}_C, \tag{28}$$

and

$$\dot{\lambda} = -\frac{2\lambda}{f_\lambda} \left\{ \mathbf{r}_{BC}^T \mathbf{Y}^{-1} \mathbf{v}_C + \left[ \mathbf{r}_{AC}^T \mathbf{Y}^{-1} \mathbf{\Gamma}_B \mathbf{r}_{BC} - \mathbf{r}_{BC}^T \mathbf{Y}^{-1} \mathbf{\Gamma}_A \mathbf{r}_{AC} \right] \mp \lambda (1-\lambda) \mathbf{n}^T \mathbf{Z} \left[ (\boldsymbol{\omega}_B - \boldsymbol{\omega}_A) \boxtimes \mathbf{n} \right] \right\}. \tag{29}$$

## III. EDMD FOR ELLIPSES AND ELLIPSOIDS

Having developed the necessary tools for dealing with overlap between ellipses and ellipsoids in Section II, we can now complete the description of the EDMD algorithm. We first discuss the fundamental step of predicting collisions between moving ellipsoids, and then explain how to process a binary collision between two ellipsoids.

### A. Predicting Collisions

The central step in event-driven MD algorithms is the prediction of the time-of-collision for two moving particles, as well as the time when a particle leaves its bounding neighborhood. This is also the most time-consuming step, especially for nonspherical particles. Although general methods can be developed for particles of arbitrary shape [43], efficiency is of primary concern to us and we prefer specialized methods which utilize the properties of ellipsoids, in particular, their smoothness and the relative simplicity of the time derivatives of the overlap potentials given in Section II D. In three dimensions, we restrict consideration to ellipsoids with a *spherically symmetric moment of inertia*, i.e., ellipsoids with equal moment of inertia around all axes. This is because the force-free motion of general ellipsoids, as well as their binary collisions, are very complex to handle. For example, the angular velocity is *not* constant but oscillates in a complex manner. It is not hard to adapt the algorithms presented here to ellipsoids with several different moments of inertia, at least in principle.

Essentially, predicting the collision time $t_c$ between two moving ellipsoids $A(t)$ and $B(t)$ consists in finding the first non-zero root of the overlap potential $F(t) = F[A(t), B(t)]$, where $F$ can be either one of $F_{AB}$, $F_B$ or $G_B$, depending on the type of collision and the choice of the potential. Formally:

$$t_c = \min t$$
$$\text{such that} \quad F(t) = 0 \quad \text{and} \quad t \geq 0, \tag{30}$$

where $F(t)$ is a smooth continuously differentiable function of time, as illustrated in Fig. 2. This kind of first root location problem has wide applications and has been studied in various disciplines. For a general non-polynomial $F(t)$, its rigorous solution is a very hard problem and requires either interval methods [6] or rigorous under/over estimation of $F(t)$ based on knowledge of exact bounds on the Lipschitz constant of $\dot{F}$ (and possibly of $F$) [10]. These methods are rather complex and are focused on robustness and generality, rather than efficiency. For particular forms of $F(t)$, rigorous algebraic methods may be possible, such as for example the prediction of time of collision of two needles (infinitely thin hard rods) [18], and possibly spherocylinders. However, this requires a considerable algebraic complexity and is not easy to adapt to a new particle shape, especially ellipsoids, for which there is not even a closed-form expression for the overlap potential.

In particular, the very elegant method for determining the time of collision of two needles proposed in Ref. [18] is related to the one proposed in Ref. [10], and at its core is the need to determine a good local or global estimate of the Lipschitz constant of $\dot{F}$ [10], i.e., an upper bound on $\left|\ddot{F}\right|$ (these are used to construct rigorous under- or over-estimators of $F(t)$). Such a global upper bound has been derived for the case of needles [c.f. Eq. (20) in [18]], but for ellipsoids the expression for $\ddot{F}$ (which we do not give here) is very complex and we have not been able to generalize the approach in Ref. [18]. As discussed in Ref. [10], significantly better results are obtained when local estimates of the Lipschitz constant of $\dot{F}$ are available (i.e., upper bounds on $\left|\ddot{F}\right|$ over a relatively short time interval), and this seems an even harder task. Nevertheless, it is a direction worth investigating in the future.

For the purpose of EDMD, it is sufficient only to ensure that an interval of overlap $[t_c, t_c + \Delta t_c]$ is not missed if

$$\min_{t_c \leq t \leq t_c + \Delta t_c} F(t) < -\epsilon_F,$$

where $\epsilon_F$ is some small tolerance, typically $10^{-4} - 10^{-3}$ in our simulations, or alternatively, if $\Delta t_c > \epsilon_t$. The use of $\epsilon_F$ is preferable because it is dimensionless with a scale of order 1. This essentially means that it is permissible to miss *grazing collisions*, i.e., collisions in which two ellipsoids overlap for a very small amount and/or for a very short time. It certainly is not productive to try to decide if two nearly touching particles are actually overlapping more accurately then the inherent numerical accuracy of $F(t)$. The choice of $\epsilon_F$ is determined by the relative importance of correctness versus speed of execution, as well as the stability of the simulation. A large $\epsilon_F$ can lead to unrecoverable errors in the event-driven algorithm, such as runaway collisions or increasing overlap between particles.

Homotopy methods can be used to solve problems such as (30). They typically trace the evolution of the root of an equation (starting from $t = 0$ in this particular case) as the equation is deformed from an initial simple form to a final form which matches $F(t) = 0$ [46]. An ordinary differential equation (ODE) solver can be used for this purpose. An essential component in these methods is *event location in ODE*s, namely, methods which solve an ODE for a certain variable $f(t)$ and determine the first time that $f(t)$ crosses zero [34, 35]. We have tested a (simple) ODE-based homotopy method for solving (30), however, since the problem at hand is one-dimensional, one can directly apply ODE event location to $f(t) \equiv F(t)$, using an absolute tolerance of $\epsilon_F$ for the ODE solver, and locate the first root $t_c$ directly more efficiently.

The ODE to solve is given by eq. (26) or (28). However, also needed is $\lambda(t)$, and one has the option of either explicitly evaluating $\lambda$ at each time step (reusing the old value of $\lambda$ as an initial guess), or also including $\lambda(t)$ in a system of ODEs using eq. (27) or (29). The second option has the advantage that one no longer needs to explicitly evaluate $\lambda$ (other then at the beginning of the integration), however, it has the additional cost of two variables instead of one in the ODE solver, which additionally leads to smaller time steps in the ODE integrator. Our numerical experiments have indicated that at least in two and three dimensions it is somewhat advantageous to explicitly evaluate $\lambda$ and only include $F(t)$ in the ODE. This may be reversed for different particle shapes, depending on the relative complexity of evaluating

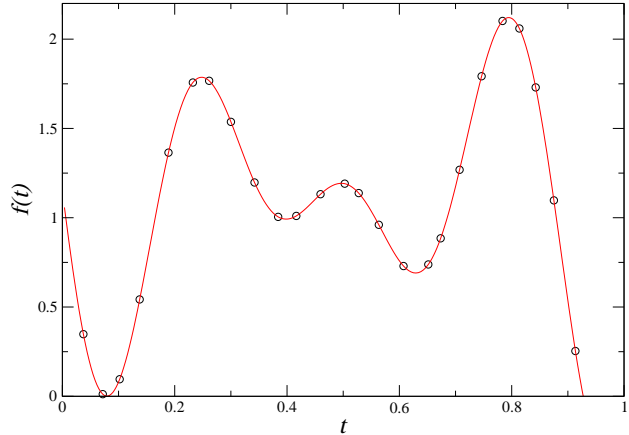

Figure 2: The time evolution of the overlap function $F_{AB}(t)$ during an ellipse collision. The overlap function is evaluated on an adaptive grid, which has a smaller time step when $F_{AB}$ changes rapidly, and a larger step when it is relatively smooth. The tracing stops when a zero crossing is detected.

$\lambda$ versus evaluating $\dot{\lambda}$.

Once $\lambda$ is evaluated however, very little extra effort is needed to evaluate $F$ explicitly, so it seems somewhat pointless to solve an ODE for $F(t)$. We have developed a method with a similar structure to ODE integration, but which uses explicit evaluation of both $F$ and $\dot{F}$. The basic idea is to take a small time step $\Delta t$, evaluate both $F$ and $\dot{F}$ at the beginning and end of the step, and use these values to form a cubic Hermite interpolant $\tilde{F}(t)$ of $F(t)$ over the interval $\Delta t$. A theoretically-supported estimate for the absolute error of the interpolant can be obtained by comparing the interpolant $\tilde{F}$ and $F$ at the midpoint of the time step, and this error can be used to adaptively increase or decrease the size of the step so as to keep the absolute error within $\epsilon_F$. This is illustrated in Fig. 2. When the interpolant crosses the $F = 0$ axes, the first root of the interpolant is used as an initial guess in a safeguarded Newton algorithm to find the exact root $t_c$. The initial time step $\Delta t$ needs to be sufficiently small to make the initial error estimate valid, and can easily be obtained by estimating a time-scale for the collision from the sizes and velocities of the particles involved in the collision. Even if this initial guess is conservative, the algorithm quickly increases the step to an appropriate value. We will refer to this algorithm as *trace event location*, since the function $F(t)$ is explicitly traced until a zero-crossing is found.

There are several details one needs to be attentive to when predicting collisions inside an EDMD algorithm. For example, some pairs of particles may already be overlapping by small amounts after having collided. In this case one can look at the sign of $\dot{F}$ to decide whether the two particles are about to have a collision or just had a collision. Two particles *can* have a collision after having collided *without* an intervening collision with third party particles. This can always happen for aspherical particles, as noted in Ref. [18], and it can also happen for spheres when boundary deformations are present (since the particles travel along curved paths), however, it cannot happen for spheres in traditional EDMD algorithms. If the initial $F$ is very close to zero, an additional safety measure is to add a small positive correction to $F$ to ensure that it is sufficiently far from zero at the beginning of the search (as compared to the accuracy in the evaluation of $F(t)$). In particular, such precautions are necessary at very high densities (i.e., near jamming).

### 1. Predicting Collisions for Ellipsoids

We sketch the procedure for predicting collisions between two moving ellipsoids in Algorithm 2. Features in Algorithm 2 include limiting the number of steps in the event location algorithm to avoid wasting resources on predicting collisions that may never happen, as well as allowing the exact prediction to fail. This algorithm first uses collision prediction for the bounding (or contained) spheres, in order to eliminate obvious cases when the particles do not collide, and to identify a short search interval for the event location by calculating the time interval during which the enclosing spheres overlap. Recall that when the boundary is not deforming this step entails solving a quadratic equation, while it involves solving a quartic equation when the lattice velocity is nonzero. In this context not only the first root of this quadratic/quartic equation needs to be determined, but also the second one, giving the interval of overlap. It may be possible to further improve the initial collision prediction step by using a bounding body other than spheres, for example, oriented bounding boxes (orthogonal parallelepipeds) [15, 21]. However, orientational degrees of freedom need to be eliminated since they are too hard to deal with because of the appearance of trigonometric functions. For example, the bounding body can be a cylinder whose axis is the axis of rotation of the ellipsoid and whose radius and length are sufficient to bound the rotating ellipsoid for all angles of rotation.

An important problem we have encountered in practice is the numerical evaluation of $G_B$ when predicting the time of collision of an ellipsoid with its bounding neighborhood.

Namely, as the particles move, it is highly likely that a point where $\mathbf{Y}$ is singular will be encountered. At such points the evaluation of $G_B$ is numerically unstable and often leads to unacceptably small time steps. We have dealt with this problem in an *ad hoc* manner, by simply trying to skip over such points, so that the search for a collision can be continued, but a more robust and systematic approach may be possible.

---

**Algorithm 2** Predict whether two moving ellipsoids overlap during the time interval $[0, T]$ by more than $\epsilon_F$, and if yes, calculate the time of collision $t_c \in [0, T]$. Essentially the same procedure can be used to determine the time a particle $A$ collides with its bounding neighborhood $B$. If the prediction cannot be verified, return a time $\tilde{t}_c < t_c$ before which a collision will not happen. Also return $\lambda$ at the time of collision if desired.

1. For all intervals of overlap of the bounding spheres of $A$ and $B$ (or for all intervals during which the bounding sphere of $A$ intersects the shell between the contained and bounding sphere of its bounding neighborhood $B$), $[t_{\text{start}}, t_{\text{end}}]$, starting from $t = 0$ and in the order of occurrence, do:

    (a) If $t_{\text{start}} > T$, return reporting that no collision can happen.

    (b) Set $t_{\text{end}} \leftarrow \min\{T, t_{\text{end}}\}$.

    (c) Update the ellipsoids to time $t_{\text{start}}$, and evaluate the initial $F$ and $\lambda$.

    (d) If $F < \epsilon$ (ellipsoids are overlapping or nearly touching), then evaluate $\dot{F}$ and:

       i. If $\dot{F} \geq 0$ then set $F \leftarrow \epsilon$ (the ellipsoids are moving apart),

       ii. Else return $t_c = t_{\text{start}}$ (the ellipsoids are approaching).

    (e) Use trace event location to obtain a good estimate of the first root of $F(t)$ during the interval $[t_{\text{start}}, t_{\text{end}}]$, putting a limit on the number of time steps (for example, in the range $100 - 250$). If no root crossing is predicted, continue with the next interval in step 1. If the search terminated prematurely, then return the last recorded time $\tilde{t}_c = t$.

    (f) Bracket the estimated first root of $F(t)$ and refine it using a safeguarded Newton's method (this may fail sometimes).

    (g) If the root refinement failed, set $t_{\text{end}} \leftarrow \frac{1}{2}(t_{\text{start}} + t_{\text{end}})$, repeat step 1c and go back to step 1e (attempt to at least find a valid $\tilde{t}_c$).

2. Return reporting that no collision will happen.

### B. Processing Binary Collisions

The steps necessary to process a binary collision between two hard particles are similar for a variety of particle shapes, and essentially involves exchanging momentum between the two particles. We give a recipe for colliding ellipsoids with a spherically symmetric moment of inertia in Algorithm 3. To determine things like the pressure it is useful to maintain the collisional contribution to the stress $\boldsymbol{\sigma}_c$ [28], which is a suitable average of the exchange of momentum over all collisions.

## IV. PERFORMANCE RESULTS

In this section we present some results for the performance of the algorithm. Many previous publications have given performance results for EDMD for spheres, and most of these results apply to our algorithm. Exact numbers depend critically on details of the coding style, programming language, compiler, architecture, etc., and are not reported here. Rather, we try to get an intuitive feeling of how to choose the various parameters of the simulation to improve the practical performance. Our main conclusion is that using NNLs is significantly more efficient than using just the cell method for particles with aspect ratio significantly different from one (greater than 2 or so) or at sufficiently high densities. Additionally, using BSCs offers significant efficiency gains for very prolate particles, for which good bounding sphere complexes can easily be constructed.

As derived in Ref. [37], when only the cell method is used, optimal complexity of the hard-sphere EDMD code is obtained when the number of cells is of the order of the number of particles, $N_c = \Theta(N)$, with asymptotic complexity $O(\log N)$ per collision, which comes from the event-heap operations. In practice however the asymptotic logarithmic complexity is not really observed, and instead to a very good approximation the computational time expanded per processed *binary collision* is constant for a given aspect ratio $\alpha$ at a given density, for a wide range of relative densities (volume fractions) $\varphi$. Even though in principle the basic EDMD algorithm remains $O(\log N)$ per collision, our aim is to improve the constants in this asymptotic form, and in particular, their dependence on the shape of the particle (in particular, the aspect ratio $\alpha$).

It is important to note that on modern serial workstations, the EDMD algorithm we have presented here is almost entirely CPU-limited, and has relatively low memory requirements, even when using NNLs and BSCs. Floating point operations dominate the computation, but memory traffic is also very important. In our implementation, simulating ellipsoids is about an order of magnitude slower than simulating spheres, even for nearly spherical ellipsoids, simply due to the high cost of the collision prediction algorithm (the same observation is reported in Ref. [18]) and increased memory traffic. For example, on a 1666 MHz Athlon running Linux, our Fortran 95 implementation uses about 0.1ms per sphere collision for a wide range of system sizes and densities. With all the improvements described in this paper, and in particular, the use of NNLs and BSCs, our implementation uses about 2ms per ellipsoid collisions for prolate spheroids, and about $2-4$ms for oblate spheroids at moderate densities for a wide range $\alpha = 1-10$. Including boundary deformations, i.e., solving quartic instead of quadratic equations when predicting binary collisions, slows down the simulation for spheres by about a factor of 2.5 (we use a general quartic solver, and better results may be obtained for a specialized solver).

## A. Tuning the NNLs

We have performed a more detailed study of the performance of the algorithm when NNLs are used, since this is a novel technique and has not been analyzed before. We perform an empirical study rather then a theoretical derivation because such a derivation is complicated by the fact that the neighborhoods evolve together with the particles, and because the numerous constant factors or terms hidden in the asymptotic expansions of the complexity actually dominate the practical performance.

Computationally, we have observed that it is good to maximize the number of cells $N_c$, even at relatively low density $\varphi \approx 0.1$, especially for rather aspherical particles. This is because binary collision predictions become much more expensive than predicting or handling transfers, and so the saving in not predicting collisions unnecessarily offsets the higher number of transfers handled. Consistent with the results reported in [37], we observe that the number of checks due to invalidated event predictions is comparable and sometimes slightly larger than the number of collisions processed, and this suggests that additional improvements in this area might increase performance noticeably.

We have tested both methods for updating the neighbor lists, the *complete* and the *partial*

update. A complete update/rebuild of the near neighbor lists after they become invalid is the traditional approach in most TDMD algorithms appearing in the literature. Since MD is usually performed on relatively homogeneous systems, when one particle displaces by a sufficient amount to protrude outside of its bounding neighborhood, most particles will have displaced a significant amount, and so rebuilding their NNLs is not so much of a waste of computational effort. The main advantage of this approach is that it can be used to build NNLs when a good estimate of $\mu_{\text{cutoff}}$ is not available, but rather a bound on the number of neighbors per particle $N_i$ is provided (this is very useful, for example, in the very early stages of the Lubachevsky-Stillinger algorithm, when particles grow very rapidly). Additionally, the algorithm for rebuilding the lists is simpler and thus more efficient. Finally, a complete rebuilding of the NNLs yields neighbor lists of higher quality, in the sense that the structure of the network of bounding neighborhoods is better adapted to the current configuration of the system and thus the size of the neighborhoods is maximal.

The algorithm for partial updates on the other hand is more complicated, and to our knowledge has not been used in MD codes. It requires using dynamic linked lists, and it will in general yield smaller average neighborhood size than a complete update, since the particle whose NNL is being updated must adjust its list without perturbing the rest of the NNLs. Note that at the beginning the NNLs must be initialized by using a complete update. The main advantage of the partial update scheme is that it is more flexible in handling nonisotropic systems or the natural fluctuations in an isotropic one. Just because one particle happened to move fast and leave its neighborhood does not mean that all particles move that fast. This is especially true at lower densities where clustering happens. In clusters particles have more collisions per unit time and thus require fewer updates of the NNLs, but outside clusters particle move large distances without collisions and thus require more frequent updates of their NNLs. In this sense partial updates are local in nature while complete updates are global. We have indeed observed that in most cases it is advantageous to use partial updates, rather then the traditional complete updates.

The first and most important test is to determine whether using near neighbor lists offers any advantages over using just the cell method. The following intuitive arguments seem clear:

- At very high densities, when the system of particles is nearly jammed [42], using NNLs

is optimal, regardless of the aspect ratio of the particles. This is because the particles move very little while they collide with nearly the same neighbor particles over and over again (see Fig. 8). Therefore near jamming the NNLs are rarely updated and by predicting the collisions only with the particles with which actual collisions happen significant savings can be obtained. However, as the density is lowered, the lists need to be updated more frequently and the complexity of using NNLs becomes significant.

- At very low densities the cell method is faster, even for very aspherical particles. This is because a particle will have many collisions with the bounding neighborhoods before it undergoes a binary collision, so that the cost is dominated by the cost of maintaining the NNLs instead of processing collisions.

- The more aspherical the particles, the more preferable the NNL method becomes compared to the cell method. This is because for very elongated particles at reasonably high densities there will be many particles per cell so using the cell method will require predicting many binary collisions that will never happen, while the NNL method will predict collisions with significantly less (truly) neighboring particles. For large $\alpha$, the dominating cost is that of rebuilding the NNLs (since this step uses the cells), and therefore the primary goal becomes to minimize the number of NNL updates per number of binary collisions processed, as well as to improve the efficiency of the NNL rebuild, i.e., using BSCs.

Our experimental results shown in Fig. 3 support all of these conclusions. We show the ratio of the CPU time expanded *per processed binary collision* for the NNL method and for the traditional cell method. We show results for equilibrium systems of prolate spheroids of aspect ratios $\alpha = 1, 3$ and $5$ at densities $\varphi = 0.1, 0.3, 0.5$ and $0.6$. Note that hard spheres jam in a disordered metastable state at around $\varphi = 0.64$, and that for the case $\alpha = 5$ we use $\varphi = 0.55$, since the jamming density is slightly lower than 0.6 for this aspect ratio [12]. In Fig. 3(a) we show the relative *slowdown* caused by using NNLs for the systems for which using the cell method is better. In Fig. 3(b) we show the relative *speedup* obtained by using NNLs for the systems for which it is better to use NNLs. Both the results of using partial and complete updates are shown.

As explained in the first part of this series of papers, we use two techniques to limit the number of neighbors that enter in the NNLs. The first one is to simply use the upper bound on the number of neighbors (interactions) $N_i$ to choose only the nearest $N_i$ neighbors per particle, and the second one is to choose a relatively small cutoff $\mu_{\text{cutoff}}$ for the maximal size of the bounding neighborhood $\mathcal{N}(i)$ (compared to the size of the particle $i$). In practice, only the second approach can be used with partial updates. This is because partial updates must work under the limitations of doing as little change to the NNLs as possible, and this requires that there be enough room to add and remove interactions from the lists as necessary. So when using partial updates one must set $\mu_{\text{cutoff}}$ to a reasonable value and then set $N_i$ to be larger then the maximal number of neighbors a particle will have given the cutoff $\mu_{\text{cutoff}}$ and an unlimited $N_i$. Reasonable values for spheres and not too aspherical particles are to set $\mu_{\text{cutoff}}$ so that on average each particle has about $5-7$ neighbors in two or $11-15$ neighbors in three dimensions (the kissing number for spheres is 6 in two and 12 in three dimensions), while setting $N_i$ at about 10 in two and 20 in three dimensions.

Since we wish to compare partial and complete updates, we change $\mu_{\text{cutoff}}$ and always set $N_i$ to a sufficiently high number (which grows sharply with $\mu_{\text{cutoff}}$), and compare partial and complete updates in Fig. 3. As expected, there is an optimal value of $\mu_{\text{cutoff}}$ which is larger for complete updates (for which NNL updates are significantly more expensive) and also at lower densities. Note however that sometimes the computation speed may change discontinuously as $\mu_{\text{cutoff}}$ is increased because at some point more than first-neighbor cells need to be searched during the NNL update. Important observations to note include the fact that *tuned partial NNL updates almost always outperform tuned complete updates*, and are thus preferred. Another useful observation is that the computation time is not very sensitive to the exact value of $\mu_{\text{cutoff}}$. Finally, note that as much as an order of magnitude of improvement is achieved for rather aspherical particles at high densities by using the NNLs. This would be even more pronounced for larger aspect ratios such as $\alpha = 10$.

For large aspect ratios, the dominant cost is that of rebuilding the NNLs, during which many particle pairs need to be tested for neighborhood. Therefore, the most important factor for the speed of the simulation is how many particles need to be examined as potential near neighbors of a given particle $i$ when rebuilding $\text{NNL}(i)$. If bounding spheres are not used, this number is proportional to the number of particles that can fit in a cell of length $\alpha$, i.e., a cube of volume $\alpha^3$. For prolate spheroids this number is proportional to $\alpha^2$, but for oblate ones it is only proportional to $\alpha$. Therefore the simulation of, for example, $\alpha = 10$, is prohibitively expensive for prolates, but not for oblates. As the results in Fig. 4 demonstrate, using

BSCs significantly increases the speed of processing collisions for very prolate spheroids. In this figure we show the approximate speedup obtained by using BSCs for tuned partial updates for a range of moderate densities and a range of aspect ratios. As expected, at large densities there are very few updates to the NNLs and therefore using BSCs does not offer a large speedup. For oblate spheroids in the same range of aspect ratios, using BSCs does not offer computational savings, and therefore we do not show any performance results. However, it is important to note that when using NNLs (at sufficiently high densities) and BSCs (at sufficiently high aspect ratios) the actual (tuned) processing time per collision is approximately the same for any ellipsoid shape in this range of aspect ratios, somewhere in the range $1 - 5$ms per binary collision in our implementation. Therefore, for practical purposes, we feel that the algorithms presented in this paper can handle a wide range of ellipsoid shapes very well.

With the use of BSCs, prolate ellipsoids are handled much better and the scaling reduced to nearly independent of $\alpha$ (note that one needs to examine $\alpha$ bounding spheres per particle, which is much less expensive than looking at neighbor particles, but still not free), as we have demonstrated above. It remains a challenge to find a technique that will also reduce the scaling to nearly independent of $\alpha$ for oblates as well. Additionally, it is important to develop a theoretical analysis of the performance of the novel steps in the algorithm, and in particular, to give estimates of the number of particles which need to be examined when building the NNLs (per particle), the number of NNL updates which need to be processed per binary collision (per particle).

### 1. Automatic Tuning of $\mu_{\text{cutoff}}$

It is important to note that it is possible to automatically tune $\mu_{\text{cutoff}}$ during the course of the simulation, at least in a rough way, so that the optimal computation speed is approached. This is very important in the Lubachevsky-Stillinger packing algorithm, since there the density is not constant but rather increases until jamming is reached. Clearly in the beginning a larger $\mu_{\text{cutoff}}$ is needed, while near the jamming point $\mu_{\text{cutoff}}$ can be set very close to 1. By monitoring the fraction of events which are collisions with a bounding neighborhood, and other statistics, one can periodically adaptively increase or decrease $\mu_{\text{cutoff}}$ during the course of the simulation. We have successfully used such techniques to speed the process of obtaining hard-particle packings, for both spheres and ellipsoids, especially for elongated ellipsoids, but do not report details here.

## V. APPLICATIONS

In this section, we present three interesting physical investigations that could not have been undertaken without the algorithm presented in this series of papers. In Section V A, we illustrate how collision-driven molecular dynamics can be used to generate tightly jammed packings of ellipsoid with densities far surpassing previously achieved ones. In Section V B, we show equation-of-state curves for quasi-equilibrium melting of an unusually dense crystal ellipsoid packing [11]. Finally, in Section V C we show how the near neighbor lists can be used to monitor the collision history near the jamming point and observe contact force chains.

### A. Generating Jammed Packings

We have used our collision-driven MD algorithm to implement a generalization of the Lubachevsky-Stillinger sphere packing algorithm [24, 25] for ellipses and ellipsoids [12]. The method is a hard-particle molecular dynamics (MD) algorithm for producing dense disordered as well as ordered packings. Small ellipsoids are randomly distributed and randomly oriented in a box with periodic boundary conditions and without any overlap. The ellipsoids are given velocities and their motion followed as they collide elastically and also expand uniformly, using the event-driven algorithm. After some time, a jammed state with a diverging collision rate and maximal density is reached. Based on our experience with spheres [22], we believe that our algorithm produces final states that are jammed, an in particular, that are *collectively jammed*. In short, a collectively jammed configuration of particles is one in which no subset of particles can simultaneously be continuously displaced so that its members move out of contact with one another and with the remainder set [13, 42]. The resulting packings are significantly denser that have previously been obtained for ellipsoids using RSA, sedimentation, or shaking (MC-like) packing protocols [4, 5, 7, 36]. High packing densities have been obtained for spherocylinders using a force-biased MC method [47], however, we believe that these packings are not truly jammed. Additionally, spherocylinders cannot be oblate or spherically asymmetric.

Several features of the molecular dynamics algorithm are necessary for the success of this packing protocol. First, provisions need to be made to allow time-dependent particle shapes, and we have explicitly included them in the treatment in Section II D. Most importantly, a very high accuracy collision resolution is necessary at very high densities, especially near the jamming point. For this reason, a time-driven approach cannot be used to generate jammed packings, and special care needs to be taken to ensure high accuracy of the overlap potentials and the time-of-collision predictions, as is done with Newton refinement in our algorithms. Finally, the use of neighbor lists significantly improves the speed of the algorithm since most computation is expended on the last stages of the algorithm when the particles are almost jammed and the use of neighbor lists is optimal (particularly combined with adaptive strategies for controlling $\mu_{\text{cutoff}}$). Including a deforming boundary in the algorithm additionally allows for a Parinello-Rahman-like adaptation of the shape of the unit cell, which leads to better (strictly jammed [13, 42]) packing of the particles (see Section V B).

We have also implemented a hard-wall spherical boundary in our algorithm, for the purpose of comparing our packings with experimental packings of MM candies and/or manufactured ellipsoids in spherical containers. The full results of these investigations will be presented in forthcoming publications [26], and here we just give an indication of the possibilities. To save time and implementation effort, we used lattice boundaries (without periodic boundary conditions) and employed a trick to implement the spherical boundary. Specifically, we put a spherical container inside a cube and then added special code in the handling of the boundary conditions to predict and process collisions with the hard walls. This was not difficult to do because the spherical container is a special case of an ellipsoid and we already have well-developed tools to deal with collisions between a small ellipsoid contained within a larger one, as presented in Section II B. In fact, implementing true flat hard walls is more difficult for ellipsoids as it necessitates the development of new overlap potentials. Figure 5 shows one of our ellipsoid packings with an unusually high density. The properties of this computer-generated packing compare very well to actual experimental data. In Fig. 6 we show how the packing density varies with the radial distance[7], clearly illustrating the layering of the particles near the hard walls and also the fact that the density inside the core of the container has a remarkable value of about 0.74, which closely matches results obtained using periodic boundary conditions and experimental results [12, 26].

---

[7] This is an approximation to the true density since it is rather nontrivial to exactly evaluate the volume of intersection of two ellipsoids.

## B.  Melting Dense Ellipsoid Crystals

Hard-particle systems are athermal, and the thermodynamic properties are solely a function of the density (volume fraction) $\phi$ [1]. It is well-known that in three dimensions hard-sphere systems have a stable low-density fluid (isotropic) phase and a stable high-density solid (face centered cubic, FCC, crystal) phase, with a first-order phase transition at intermediate densities [41]. Determining the exact transition densities is rather difficult and requires evaluating free energies via thermodynamic integration [19]. Nevertheless, the first-order phase transitions can be directly observed in molecular dynamics simulations, and the relevant dynamics (nucleation or relaxation) studied. In MD, one usually studies equilibrium properties by starting with a nonequilibrium system at a given density and then allowing it to equilibrate for a sufficiently long time. An alternative is to very slowly change the density in a quasi-equilibrium manner while tracking the relevant properties such as pressure or order-parameters. This kind of procedure allows one to directly observe the process of melting of the high-density crystal or the freezing of a liquid, and identify approximately the transition points. The collision-driven MD algorithm we have presented is ideally suited for such a study. Namely, by imposing a very small rate of expansion/contraction $\gamma$ of the particle extents, one can continuously change the density while keeping the system in quasi-equilibrium.

By starting with a low-density isotropic fluid and very slowly increasing the density, one can produce a superdense liquid and then observe a first-order freezing transition as soon as the metastable fluid becomes unstable, typically when the density approaches the maximal density of coexistence (as also observed in Ref. [40]). This freezing is a nucleation-activated (rare-event) process (the dynamics of which can be observed) and does not lead to perfect crystallization, but is clearly visible as a discontinuous pressure drop, as illustrated for hard-spheres in the inset in Fig. 7. One can reverse the process by starting with a perfect crystal, assuming that the stable high-density crystal structure is known, and slowly reduce the density until the crystal melts, typically as the density approaches the minimal density of coexistence, as illustrated in the inset in Fig. 7. In the figure, comparison is made to the semi-empirical results for the liquid, solid, and coexistence regions for hard spheres in the literature [40], and for ellipsoids comparison is made to scaled-particle theory for the isotropic fluid [38]. Unfortunately, direct coexistence is very difficult to observe in computer simulations, and requires the creation of an artificial interface between the two phases [27]. Additionally, it is in principle important to include unit cell dynamics in order to allow for solid-solid transitions. However, it is difficult to do this in a stable manner across a range of densities and in this study we fix the unit cell.

For hard ellipsoids, it is not known what is the high-density crystal structure. One can hope to identify candidate structures by densifying a liquid sufficiently slowly to allow for nucleation. By using the collision-driven algorithm with a very slow expansion and small systems (6-16 particles), and with a deforming boundary, we were able to identify crystal packings of ellipsoids that were significantly denser ($\phi = 0.771$) than the previously assumed crystal structure [11], namely, an affine deformation of the hard-sphere crystal (FCC, $\phi = 0.741$) [1]. It is important to note that this discovery was made possible because of the inclusion of boundary deformation into the algorithm, which allowed to sample a wide range of crystal structures. In Fig. 7, we show the melting of this newly-discovered two-layered ellipsoid crystal for an aspect ratio of $\sqrt{3}$ for prolate and $1/\sqrt{3}$ for oblate spheroids.

The crystal melts into an isotropic fluid and no nematic phase is observed, as can be seen by monitoring the nematic order parameters, which rapidly goes to zero as the first-order transitions occur. For comparison, we also show the corresponding melting curves for the ellipsoid crystal obtained by affinely stretching or compressing an FCC sphere crystal along the $(0, 0, 1)$ direction. Additionally, we try to observe the freezing of the isotropic liquid by slow compression. However, it can be seen that despite the very slow compression the liquid does not freeze but rather jams in a metastable glass. This illustrates that systems of ellipsoids have a marked propensity toward (orientationally) disordered configurations and are very hard to crystallize. This is to be contrasted with the case of hard spheres where we easily observe freezing at the same expansion rate. It is interesting to note that all of the pressure curves have a marked linear behavior around the jamming density $\phi_J$ when plotted with a reciprocal pressure axes, i.e., $P \sim (\phi - \phi_J)^{-1}$, in agreement with free-volume theory [32]. A close agreement between the results for prolate and oblate spheroids is seen.

Many questions concerning the stable and metastable phases for ellipsoids as a function of density and aspect ratio(s) remain open, and the algorithm developed in this paper provides a powerful tool for future studies. It is important to point out that one can use MD to locate the exact point of coexistence by calculating the free energy (i.e., the entropy for hard-particle systems) by integrating the equation of state along a reversible path starting from a configuration of known free energy, for both the fluid and the solid phases. In particular, for the solid phase one needs to use special "single-occupancy-cell" constrained MD, in which each particle is constrained to remain in a cell centered around its position in the the close-packed configuration [48]. One can in fact use the bounding neighborhood $\mathcal{N}(i)$ as the cell for particle $i$, as was done for spheres in the so-called tether method for calculating the entropy [39]. Only a minor modification to the EDMD algorithm we presented is needed, namely, when a particle collides with its bounding neighborhood it should bounce back elastically rather then rebuild its NNL, and the total pressure on the imaginary walls of the bounding neighborhoods accumulated. Such investigations will be carried out in the future.

### C. Observing Contact Force Networks

In this series of papers, we have mostly focused on using near neighbor lists as a tool to improve the efficiency of the collision-driven algorithm. However, using neighbor lists has additional advantages as well. The most important one is that it allows one to monitor the collision history of each particle or a pair of particles. This is especially useful for dense hard-particle packings near the jamming point. In the very limit of a jammed packing, each particle has a certain number of contacting geometric neighbors, and cannot displace from its current position [13]. The network of interparticle contacts forms the *contact network* of the packing, and this contact network can carry positive contact forces. For packings of soft spheres, interacting with a differentiable potential, it is easy to obtain contact forces near a jamming point and observe the resulting force chains and the distribution of contact forces, which has been noted to have an exponential tail in a variety of models of granular materials [3].

At first glance it may seem like force networks cannot be observed for hard particles since forces are ill-defined. Importantly, however, we now demonstrate that this is not true. As the jamming point is approached, for example, via the Lubachevsky-Stillinger packing algorithm, each particle collides repeatedly only with a small set of neighbors (which become contacts in the jamming limit). By averaging the total exchanged momentum between each pair of recently collided particles, one obtains a measure of the contact force between pairs of particles (with some arbitrary proportionality constant). The resulting force network can be monitored and recorded by tracking additional information for each interaction in the NNLs, such as total number of collisions between the given pair of particles, total momentum exchanged, etc. The resulting force chains in two dimensions are illustrated for a binary disk packing in Fig. 8, to be compared with similar pictures in, for example, Ref. [3]. This force network turns out to be almost in equilibrium, i.e., the total force (and torque for ellipsoids) on each particle approaches zero, and this *self stress* [8] of the packing is in a sense what causes (or certifies) the jamming of the particles [8]. The distribution of contact forces in a three-dimensional hard-sphere packing is shown in Fig. 9, and matches similar results for systems of soft particles reported in the literature.

## VI. CONCLUSIONS

In this series of two papers, we have presented a practical event-driven molecular dynamics algorithm for systems of ellipses and ellipsoids in some detail. The algorithm utilizes a number of traditional techniques, and introduces a novel use of near-neighbor lists via the concept of bounding neighborhoods. Furthermore, using bounding sphere complexes in addition to neighbor lists significantly improves the handling of very aspherical particles. We have proposed a general method for determining the time of collision of two particles of any shape for which a smooth overlap potential can be constructed and easily differentiated. The application to ellipses or ellipsoids has been developed in detail.

We have identified a number of important directions for future investigation, which can lead to significantly faster algorithms for very aspherical particles. First, predicting the time of collision of two moving particles can be improved, either by using techniques other than bounding spheres in order to narrow the search intervals during which a collision may happen, or by improving the algorithm to search those intervals for a collision. In particular, a new overlap potential for the case of one small ellipsoid contained within another large ellipsoid is needed. The practical handling of bounding sphere complexes can further be improved. More importantly, it is an open challenge to develop an algorithm to improve the efficiency of building the near-neighbor lists for very oblate particles.

The three physical applications of the algorithm that we presented could not have been undertaken with the same efficiency and accuracy without a collision-driven MD algorithm. Our packing algorithm is unique in its ability to produce hard-particle packings which are collectively jammed to a very high degree of accuracy (particle contacts reproduced to within an accuracy of $10^{-12}$) and compare well to experimental packings. The packings produced by the algorithm are significantly more dense than those produced by previously reported algorithms, which do not produce jammed packings and are less efficient. Our studies have also enabled the discovery of the densest known ellipsoid crystals, bringing into question previously reported phase diagrams for systems of ellipsoids in the high-density region. The event-driven approach can also be used to effectively study the thermodynamic properties of ellipsoid packings, and, in particular, identify the equilibrium phases and observe the dynamics of melting, freezing, and metastable liquids, at very high densities, i.e., close to the relevant jamming point, be it a close-packed crystal, or a metastable glass. Finally, we showed that one can gather collision statistics during the algorithm to obtain force chains in true hard-particle packings, which have only been reported for soft particles.

**ACKNOWLEDGMENTS**

A. D. would like to thank H. Sigurgeirsson for his help with implementing the EDMD event loop, L. F. Shampine and I. Gladwell for their advice concerning event location in ODEs, and the `freeglut` library team for their help with visualization, as well as many others who have contributed public domain codes used in this project. The authors were supported in part by the Petroleum Research Fund under Grant No. 36967-AC9, and by the National Science Foundation under Grant Nos. DMR-0213706 and DMS-0312067.

**Algorithm 3** Process a binary collision between ellipsoids $i$ and $(j, v)$. Assume that the particles have already been updated to the current time $t$, and that $\lambda$ at the point of collision is supplied (i.e., it has been stored for particle $i$, and also $1 - \lambda$ in particle $j$, when this collision was predicted).

1. Calculate the Euclidean positions and velocities of particles $\mathbf{r}_A$, $\mathbf{r}_B$, $\mathbf{v}_A$ and $\mathbf{v}_B$, as well as their angular velocities $\boldsymbol{\omega}_A$ and $\boldsymbol{\omega}_B$.

2. Find the contact point $\mathbf{r}_C$ and normal velocity at the point of contact $v_n = \hat{\mathbf{n}}^T \mathbf{v}_C$, using the supplied $\lambda$.

3. If $v_n > 0$, then return without further processing this (most likely grazing) mis-predicted collision.

4. Calculate the exchange of momentum between the particles $\Delta \mathbf{p}_{AB} = \Delta p_{AB} \hat{\mathbf{n}}$,

$$\Delta p_{AB} = 2v_n \left( \frac{1}{m_A} + \frac{1}{m_B} + \frac{\|\mathbf{r}_{CA} \times \hat{\mathbf{n}}\|}{I_A} + \frac{\|\mathbf{r}_{CB} \times \hat{\mathbf{n}}\|}{I_B} \right)^{-1},$$

where $m$ denotes mass and $I$ moment of inertia.

5. Calculate the new Euclidean velocities of the particles,

$$\mathbf{v}_A \leftarrow \mathbf{v}_A - \frac{\Delta p_{AB}}{m_A} \hat{\mathbf{n}}$$
$$\mathbf{v}_B \leftarrow \mathbf{v}_B + \frac{\Delta p_{AB}}{m_B} \hat{\mathbf{n}},$$

as well as the new angular velocities

$$\boldsymbol{\omega}_A \leftarrow \boldsymbol{\omega}_A - \frac{\Delta p_{AB}}{I_A} (\mathbf{r}_{CA} \times \hat{\mathbf{n}})$$
$$\boldsymbol{\omega}_B \leftarrow \boldsymbol{\omega}_B + \frac{\Delta p_{AB}}{I_B} (\mathbf{r}_{CB} \times \hat{\mathbf{n}}).$$

Optionally update any averages that may need to be maintained (such as average kinetic energy) to reflect the change in the velocities.

6. Record the collisional stress contribution

$$\boldsymbol{\sigma}_c \leftarrow \boldsymbol{\sigma}_c - \Delta p_{AB} \left( \mathbf{r}_{AB} \hat{\mathbf{n}}^T \right).$$

7. If using NNLs, record information about the collision that is being collected for the interaction between $i$ and $(j, v)$, such as an accumulation of the total exchanged momentum for this interaction, total number of collisions for this interaction, etc.

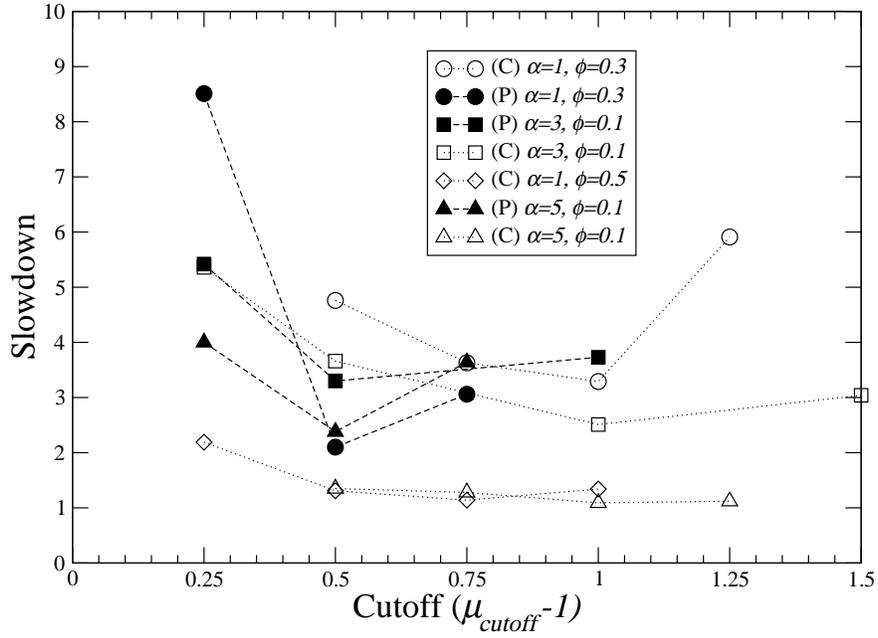

(a) Cell method wins

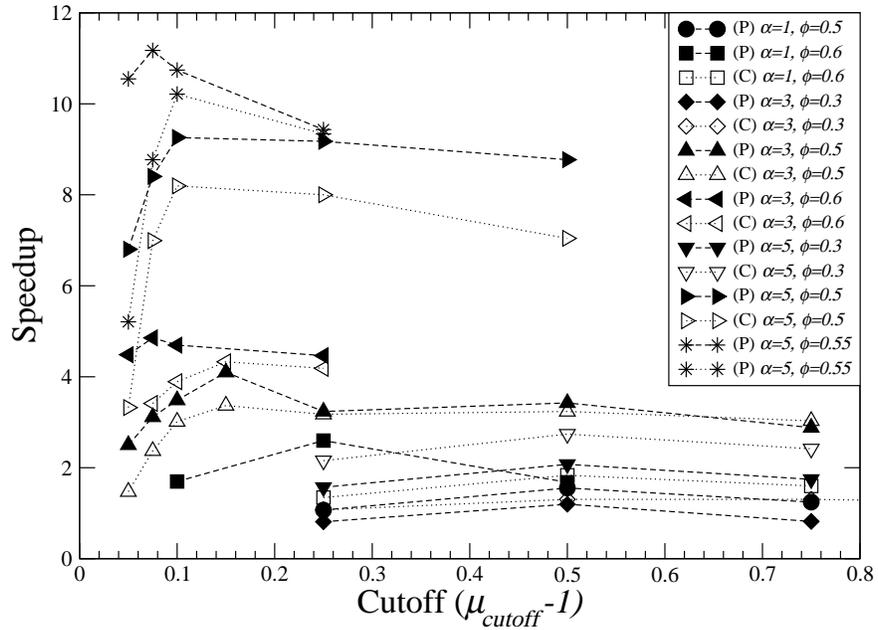

(b) NNL method wins

Figure 3: Performance results for using NNLs in addition to the traditional cell method for a variety of aspect ratios $\alpha$ and densities $\varphi$ for prolate spheroids, with both partial (P) or complete (C) updates.

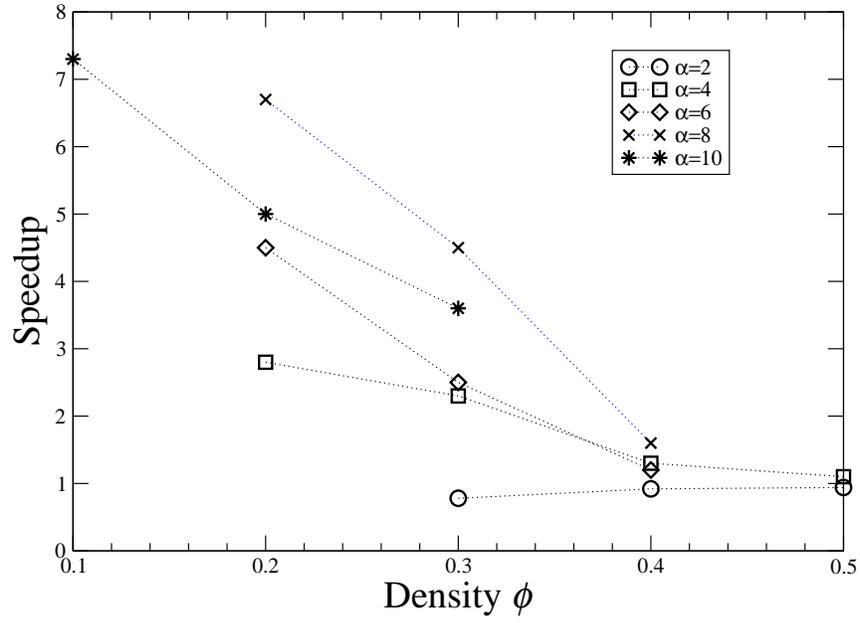

Figure 4: Performance results for using BSCs for prolate spheroids at low to moderate densities. Using BSCs does not appear to offer computational savings for oblate spheroids in the same $\alpha$ range.

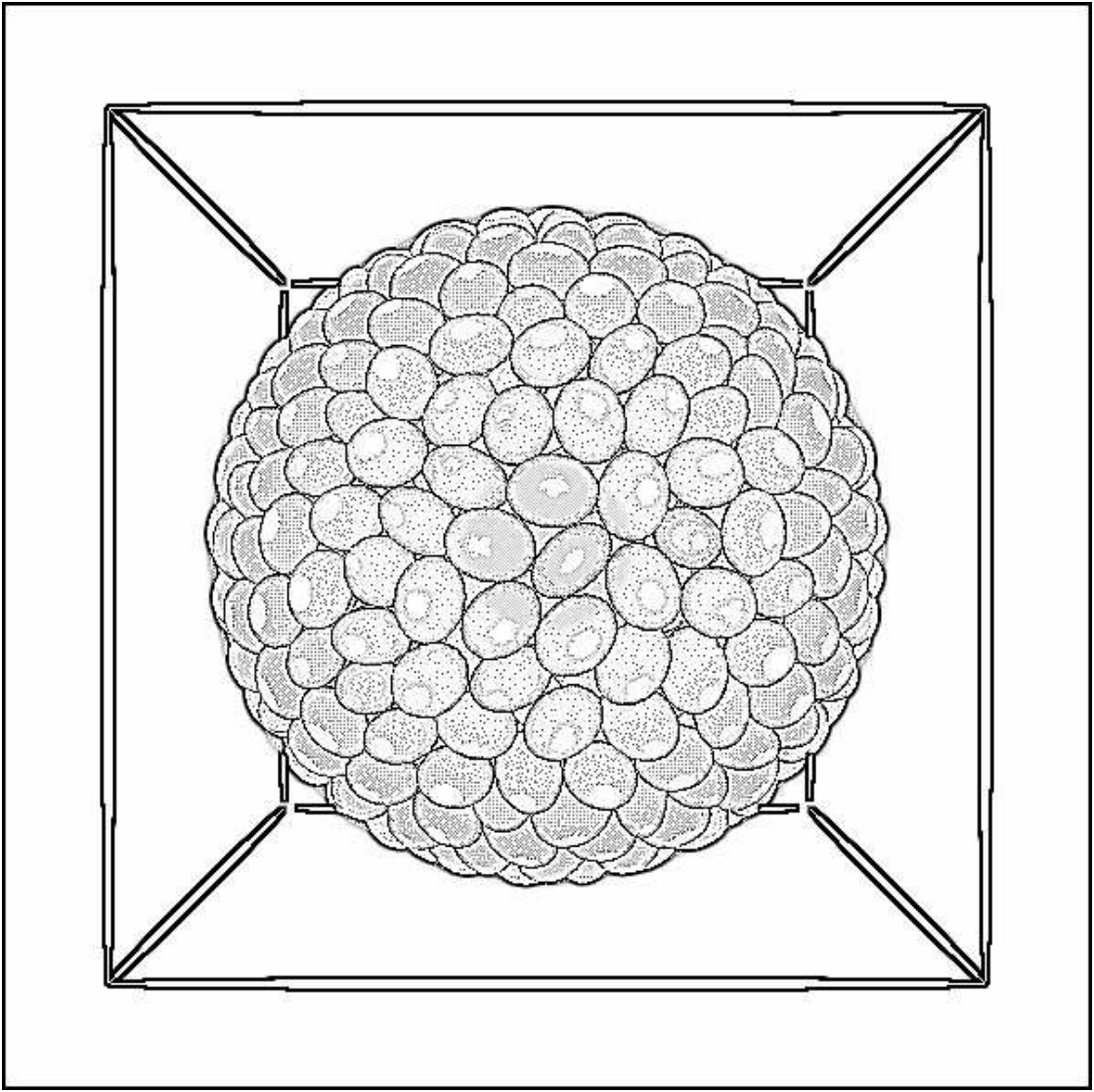

Figure 5: A packing of $N = 1,000$ ellipsoids with aspect ratios $0.8 : 1 : 1.25$ inside a spherical container. A cube enclosing the sphere is used as a pseudo-boundary for the purposes of the cell method. The radial density profile of a larger packing is shown in Fig. 6.

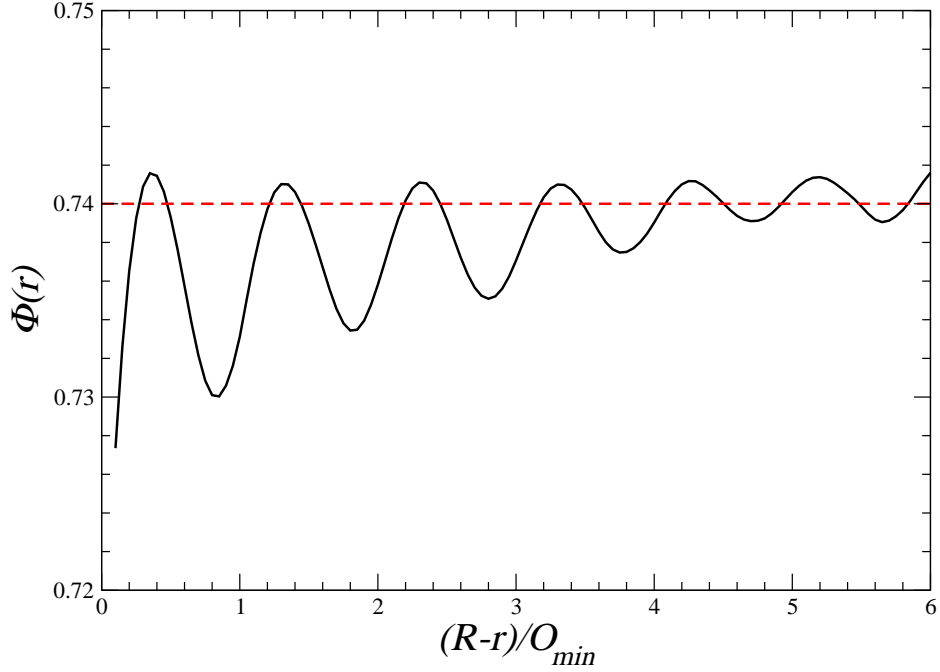

Figure 6: The (approximate) fraction $\Phi(r)$ of a sphere of radius $r$ (measured in units of the smallest ellipsoid axes $O_{\min}$) covered by the particles for a packing of $N = 5,000$ ellipsoids like those in Fig. 5 inside a spherical container of radius $R$. Due to the lower density and ordering of the particles next to the hard wall, one needs to eliminate several layers of particles close to the wall before $\Phi(r)$ reaches the core density of about $\Phi = 0.74$ (dashed line), surprisingly close to the highest possible density for sphere packings. A simulation with periodic boundary conditions gives a bulk density of about $\phi_{\text{bulk}} = 0.735$.

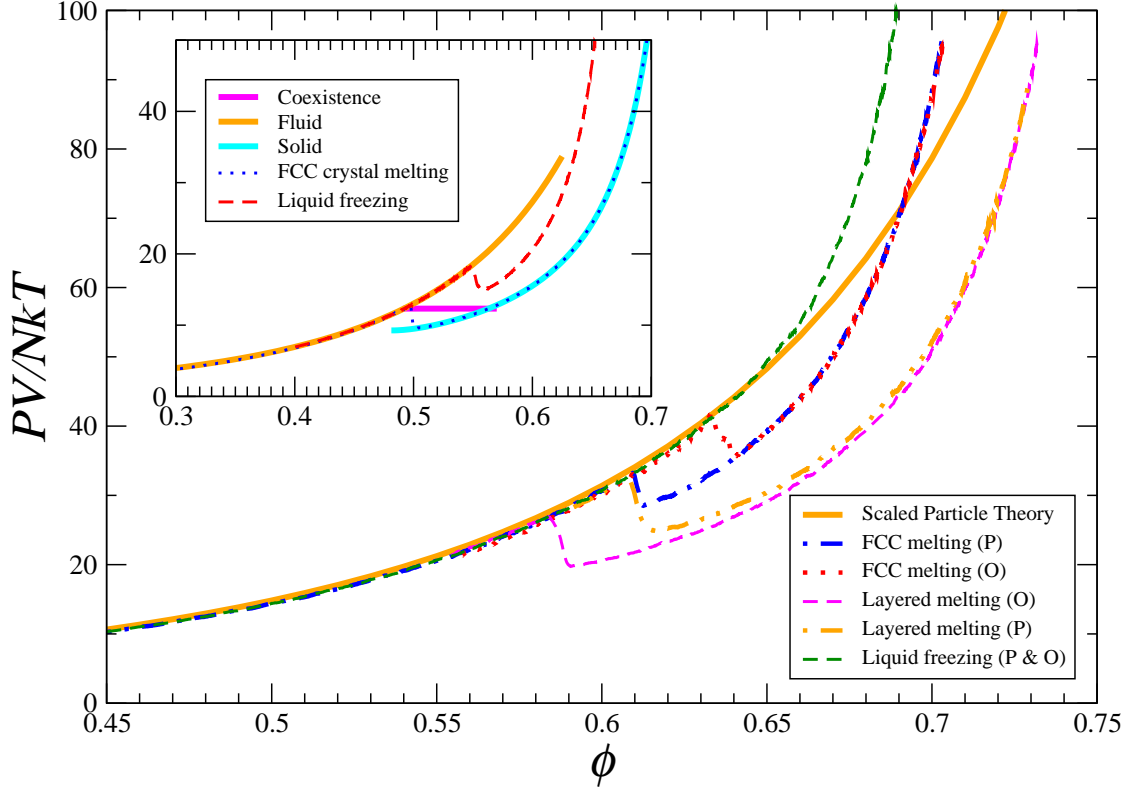

Figure 7: Equation-of-state (pressure-density) curves for prolate and oblate ellipsoids of aspect ratios $\sqrt{3}$ and $1/\sqrt{3}$ respectively, compared with spheres (inset), as obtained from quasi-equilibrium collision-driven MD with $N = 1024$ particles. The temperature $kT$ is maintained at unity by frequent velocity rescaling and the "instantaneous" pressure and order parameters are averaged and recorded every 100 collisions per particle. The unit cell is kept fixed. (A) For hard spheres, a perfect FCC crystal is slowly melted by reducing the density ($\gamma = -10^{-6}$) starting from $\phi = 0.7$, and an isotropic fluid is frozen by slowly compressed ($\gamma = 10^{-6}$) starting from $\phi = 0.4$. Very similar curves are obtained for both smaller $|\gamma|$ and for larger systems (we have used up to $N = 10,000$ particles), indicating that the observed curves are not dominated by finite-size or dynamical effects. (B) For hard ellipsoids, a very dense two-layered crystal [11] is melted from a density of $\phi = 0.73$, and similarly an affinely deformed FCC crystal is melted starting from $\phi = 0.7$ ($\gamma = -10^{-6}$). An attempt to freeze an isotropic liquid on the other hand fails and leads to a jammed metastable glassy configuration with $\phi \approx 0.72$ despite the slow expansion ($\gamma = 10^{-6}$), for both oblate and prolate ellipsoids. Investigating even slower expansion or larger systems is computationally prohibitive at present, and hence the results should be interpreted with caution.

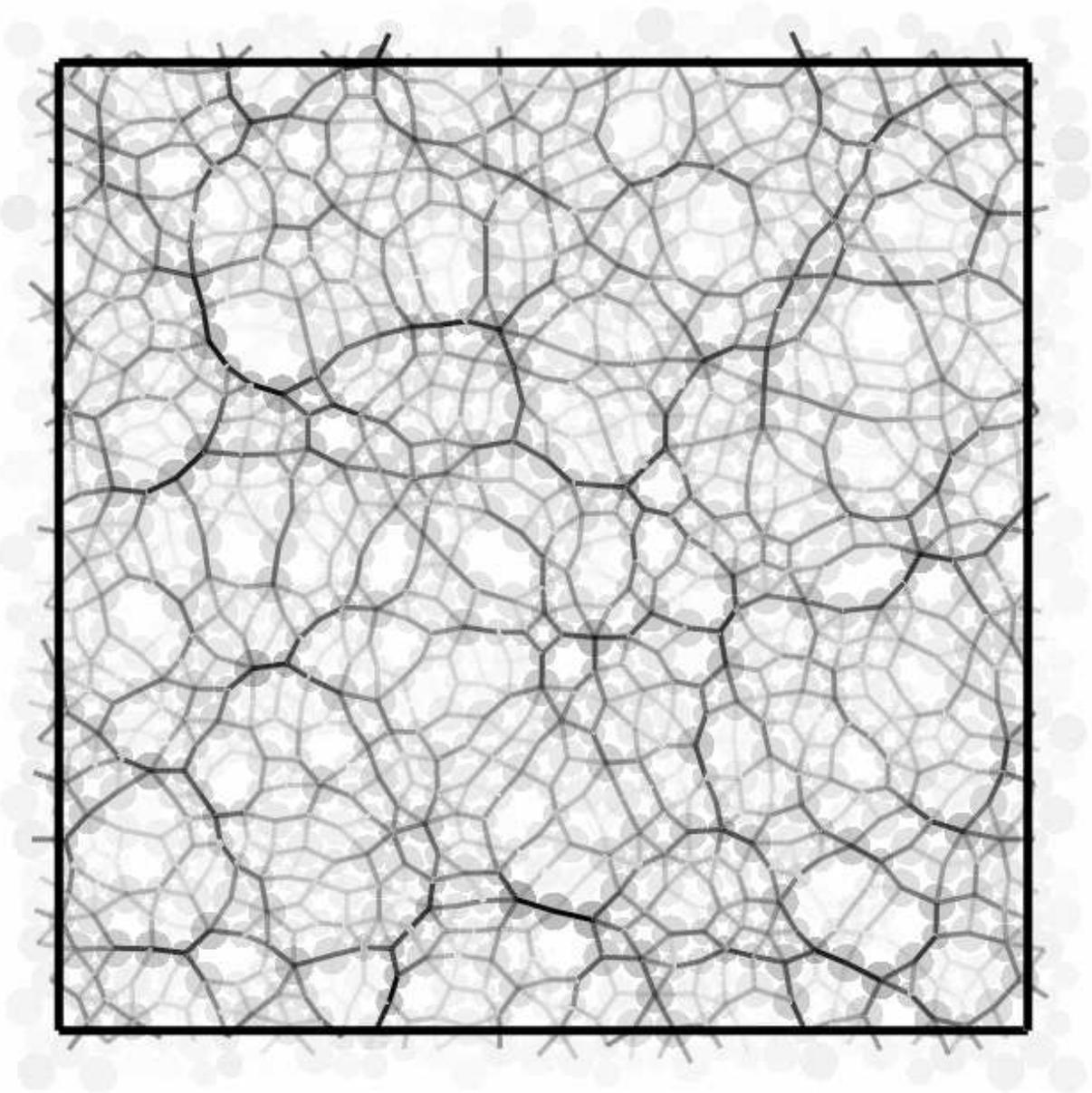

Figure 8: A network of interparticle forces (darker lines indicate stronger forces) in a binary disk packing, as obtained by averaging the total exchanged momentum between colliding particles over a long period of time during the final stages of the Lubachevsky-Stillinger packing algorithm. Darker particles collide more frequently then lighter ones.

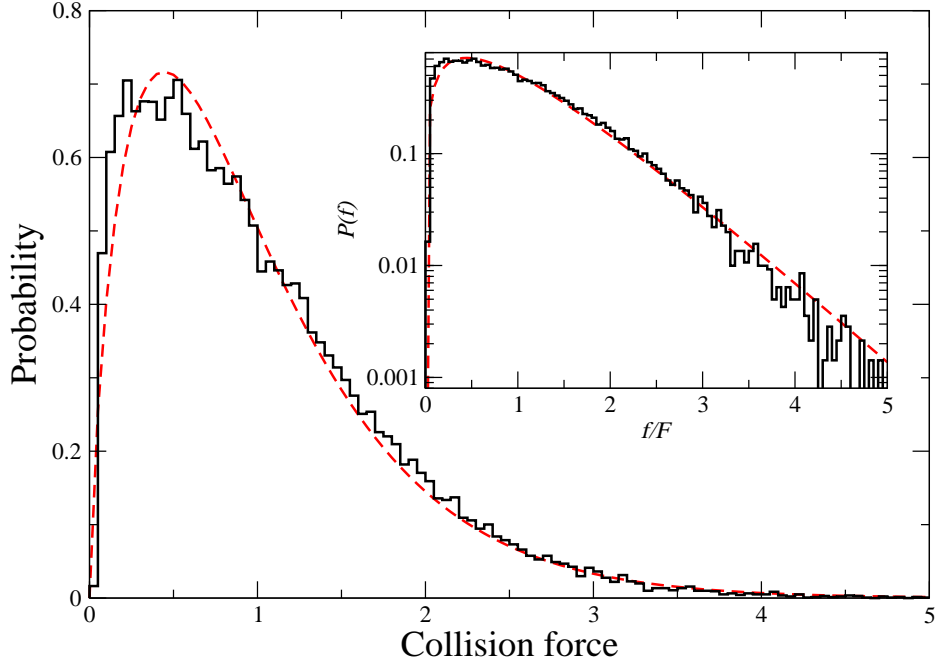

Figure 9: The probability distribution for contact forces in an almost jammed monodisperse three-dimensional hard sphere packing ($N = 10,000$ particles), along with a fit of the form $P(f) \sim f^{0.8} e^{-1.8 f/F}$, where $F$ is the mean force [3]. Only a single configuration was used, with density $\phi = 0.6438$, and the collisional forces (total exchanged momentum) were averaged over $10,000$ collisions per particle. The inset shows $P(f)$ on a logarithmic scale to emphasize the exponential tail for large contact forces.